\documentclass[a4paper,11pt]{article}
\pdfoutput=1
\usepackage{jcappub}
\usepackage[T1]{fontenc}
\usepackage{comment}
\usepackage{xcolor}
\usepackage[normalem]{ulem}
\newcommand{\ste}[1]{{\color{teal} {#1}}}

\title{\boldmath Cosmology with a Non-minimally Coupled Dark Matter Fluid\\ II. Cosmological Perturbations}

\author[a,b,c]{Samuele Silveravalle,}
\author[a,b,c,d]{Andrea Lapi,}
\author[a,b,c]{Francesco Benetti,}
\author[a,b,c]{Stefano Liberati}

\affiliation[a]{SISSA - International School for Advanced Studies, Via Bonomea 265, 34136 Trieste, Italy}
\affiliation[b]{INFN, Sezione di Trieste, Via Valerio 2, 34127 Trieste, Italy}
\affiliation[c]{IFPU - Institute for Fundamental Physics of the Universe, Via Beirut 2, 34151 Trieste, Italy}
\affiliation[d]{INAF, IRA - Istituto di Radioastronomia, Via Gobetti 101, 40129 Bologna, Italy}

\emailAdd{fbenetti@sissa.it, lapi@sissa.it, ssilvera@sissa.it, liberati@sissa.it}

\abstract{
We extend our study of a cosmological scenario in which dark matter is non-minimally coupled to gravity at the fluid level. In previous work, we showed that this interaction can drive an early phase of accelerated expansion, addressing the horizon and flatness problems, and can also lead to a cosmological bounce in the presence of negative spatial curvature. Here we analyse the evolution of linear perturbations in this framework. We derive the equations governing scalar, vector and tensor perturbations, and obtain analytic solutions in the relevant cosmological regimes. We find that perturbations generated during the accelerated expansion phase produce a strongly blue scalar power spectrum and are therefore incompatible with observations. By contrast, if primordial fluctuations are generated during the contracting phase of a bouncing solution, the model yields an approximately scale-invariant scalar power spectrum while keeping the tensor-to-scalar ratio compatible with current bounds. Although our treatment relies on simplifying approximations that should be refined in future work, these results indicate that an observationally viable primordial spectrum can be obtained only with a bouncing solution, and hence with a non-singular early-time evolution.
}

\begin{document}
\maketitle
\flushbottom

\section{Introduction}\label{sec|intro}

Our current cosmological model ($\Lambda$CDM), in which the Universe is mainly filled by a constant energy density ($\Lambda$) and by non-interacting, non-relativistic cold dark matter (CDM), has been increasingly challenged in recent years. Despite its success in reproducing a wide range of observations~\cite{SupernovaSearchTeam:1998fmf,Scolnic:2021amr,WMAP:2012fli,ACT:2020gnv,Planck2018,SPT-3G:2021eoc,SupernovaCosmologyProject:1998vns,Brout:2022vxf,DES:2024jxu,BOSS:2014hhw,BOSS:2016wmc,eBOSS:2020yzd,DESI:2024mwx,WiggleZ:2012sek,Beutler:2012px,Okumura:2015lvp,Pezzotta:2016gbo,Said:2020epb,Turner:2022gvw,White:1993wm,SPT:2021vsu,Ghirardini:2024yni}, tensions have emerged in the values of the Hubble parameter $H_0$ and the density fluctuations amplitude $\sigma_8$, when measured at late times and when inferred from cosmic microwave background (CMB) observations~\cite{DiValentino:2021izs,Riess:2016jrr,Riess:2024vfa,Heymans:2013fya,DiValentino:2020vvd,DES:2021vln}. Moreover, data from \texttt{DES} and \texttt{DESI} suggest a dark energy component with a time-dependent equation of state~\cite{DESI:2024mwx,DES:2024jxu}, in contrast with the assumption of a cosmological constant.

Within this apparent $\Lambda$CDM crisis, the fact that about $84\%$ of matter is in the form of cold dark matter remains a robust result, supported by both cosmological and astrophysical evidence~\cite{Allen2011,Rubin1980,Persic1996,Bennet2003,Aver2015,Planck2018,eBOSS:2020yzd,Pan-STARRS1:2017jku,Garrel2022,SPT:2021vsu,Clowe2006,Paraficz2016}. Nonetheless, the physical nature of dark matter remains one of the main open questions in physics, with candidates ranging from ultralight particles with masses as small as $10^{-21}\mathrm{eV}$, such as axions or axion-like particles~\cite{Duffy:2009ig,Chadha-Day:2021szb}, to compact objects with masses as large as $10^{-11}M_{\odot}\sim10^{55}\mathrm{eV}$, such as primordial black holes~\cite{Carr:2016drx,Green:2024bam}. This broad range of possibilities reflects the fact that dark matter does not interact electromagnetically by definition, and direct searches for possible weak-scale interactions have so far found no evidence~\cite{XENON1T,PANDAX,LZ}. Recent results from the LUX-ZEPLIN experiment, however, have reported a possible hint of a weakly interacting massive particle with a mass larger than $200$~GeV~\cite{LZ:2026axp}.

The elusive nature of dark matter, however, opens the possibility of probing modified gravitational interactions through its dynamics. Direct couplings between matter and curvature, generally referred to as non-minimal couplings (NMCs), arise naturally as high-energy corrections in quantum field theory in curved spacetime~\cite{Birrell:1982ix,Hertzberg:2010dc}. While these interactions are subdominant with respect to Standard Model forces for baryonic matter, they could represent the leading direct interaction for dark matter. Similar couplings may also emerge effectively on large scales if dark matter undergoes a form of Bose–Einstein condensation~\cite{Bettoni14}. To account for these different possibilities, while remaining agnostic about the microscopic nature of dark matter, we consider a non-minimal coupling acting at the fluid level.

A series of papers has investigated the properties of non-minimally coupled fluids~\cite{Bettoni11,Bettoni12,Bettoni:2013diz,Bettoni14,Bettoni15}, together with the consequences of their Newtonian limit on cosmological and galactic scales~\cite{Gandolfi21,Gandolfi22,Gandolfi23}. In two recent works, we presented a fully relativistic realization of this model and began exploring its phenomenology at both astrophysical~\cite{Benetti_2025} and cosmological~\cite{Silveravalle:2025yij} scales. In those papers, we argued that the non-minimal coupling becomes relevant whenever the mean free path of the fluid, $\lambda_{\rm mfp}=m/(\rho\sigma)$, where $\sigma$ is the total cross section, becomes smaller than the spacetime curvature radius, $R_c=c(8\pi G\rho)^{-1/2}$, namely when the density exceeds the threshold
\begin{equation}
    \rho>\rho_{\rm th}=\frac{8\pi G}{c^2}\left(\frac{\sigma}{m}\right)^{-2}.
\end{equation}
For a cross section of the order of the smallest measured in Standard Model experiments, $\sigma/m\approx 10^{-39}\mathrm{cm}^2/\mathrm{GeV}$~\cite{Kretzschmar:2018ntq}, this corresponds to a threshold density $\rho_{\rm th}\approx 10^4\mathrm{g/cm}^3$. We showed that this allows for ultra-compact self-gravitating dark matter objects with masses of order $10^5M_\odot$~\cite{Benetti_2025}, and leads to a modified cosmological evolution at redshifts larger than $z\approx 10^{11}$~\cite{Silveravalle:2025yij}.

In particular, in the paper preceding the present one~\cite{Silveravalle:2025yij}, we investigated the background evolution of a homogeneous and isotropic Universe containing a non-minimally coupled dark matter fluid. We showed that, at redshifts larger than $z\approx 10^{11}$, the Universe was in a phase of accelerated expansion. At even earlier times, in the presence of non-zero spatial curvature, the Universe either enters a ballistic expansion phase or undergoes a bounce from a previous contracting phase, if spatial curvature is positive or negative, respectively. For values of the spatial curvature compatible with current constraints, this occurs at redshifts larger than $z\approx 10^{32}$, so that the accelerated phase lasts long enough to solve the horizon and flatness problems. 

These appealing results at the background level motivate a careful analysis of cosmological perturbations, both to assess the physical viability of the model and to determine whether it can provide a genuine alternative to standard inflationary scenarios. At present, the primordial power spectrum is tightly constrained by CMB observations, in particular by Planck~\cite{Planck2018}, which indicate an almost scale-invariant spectrum of scalar perturbations and a negligible amplitude of primordial gravitational waves. Our model, however, differs substantially from standard inflation, which can usually be recast as a scalar field slowly rolling down a nearly flat potential. A dedicated analysis of perturbations is therefore required.

In section~\ref{sec|cosmo} we review the non-minimally coupled dark matter fluid model and its background cosmological evolution. In section~\ref{sec|theory} we derive the scalar, vector, and tensor perturbation equations and their solutions in the relevant regimes. Finally, in section~\ref{sec|evolution} we track the evolution of perturbations and compute the corresponding observables, namely the scalar spectral index and the tensor-to-scalar ratio.

In the main text, small capital subscripts with one or two letters refer to specific matter components, in particular we use ${}_{\textsc{b}}$ for baryons, ${}_{\textsc{dm}}$ for dark matter, ${}_{\textsc{r}}$ for radiation, and ${}_{\Lambda}$ for the cosmological constant. Subscripts with three letters refer to different cosmological epochs, in particular we use ${}_{\textsc{mat}}$ for matter domination, ${}_{\textsc{rad}}$ for radiation domination, ${}_{\textsc{nmc}}$ for non-minimal coupling domination, ${}_{\textsc{mre}}$ for the matter-radiation equality, and ${}_{\textsc{bnc}}$ for the bounce.

\section{Non-minimally coupled cosmology}\label{sec|cosmo}

In this section we briefly summarize the theoretical model for the dynamics of non-minimally coupled dark matter, and the cosmological evolution at the background level illustrated in~\cite{Silveravalle:2025yij}.

We consider the action for non-minimally coupled dark matter
\begin{equation}\label{eq|action}
   S_{\textsc{nmc}} = S_{\textsc{mc}} + S_{\textsc{int}} = \int_{\mathcal{M}}{\rm d}^4x\; \sqrt{-g}\,\left[\frac{c^4}{16\pi\,G}R + \mathcal{L}_{\textsc{sm}} + \mathcal{L}_{\textsc{dm}}\right]+\int_{\mathcal{M}}{\rm d}^4x\; \sqrt{-g}\,\epsilon L^2 G_{\mu\nu}\,T^{\mu\nu}_{\textsc{dm}},
\end{equation}
where $\mathcal{L}_{\textsc{sm}}$ denotes the Lagrangian for all non-dark matter components, $\mathcal{L}_{\textsc{dm}}$ the Lagrangian for dark matter, and the dark matter stress-energy tensor $T^{\mu\nu}_{\textsc{dm}}$ in the interaction term is that of pressureless dust: $T^{\mu\nu}_{\textsc{dm}} = \rho_{\textsc{dm}} \,c^2\,u^{\mu}\,u^{\nu}$, where $\rho_{\textsc{dm}}$ is the energy density and $u^{\mu}$ the four-velocity.

The interaction Lagrangian depends on two parameters: a characteristic length scale $L$, introduced for dimensional consistency and for setting the interaction scale, and a dimensionless polarity parameter $\epsilon=\pm 1$, which determines whether the interaction is respectively attractive or repulsive. In both cosmological~\cite{Gandolfi21} and astrophysical~\cite{Benetti_2025} contexts, only the choice $\epsilon=-1$ yields regular solutions. Concerning the length scale, we have previously determined $L \approx 3.7\times10^{11}\,\mathrm{cm}$, corresponding to masses of order $10^{5}\,M_{\odot}$ for ultra-compact objects in the astrophysical case, and to negligible effects of the non-minimal coupling after neutrino decoupling in the cosmological scenario. For consistency with our previous work, we adopt $L = 3.7\times10^{11}\,\mathrm{cm}$ as the fiducial value, set $\epsilon=-1$, and use units with $c=1$.

The equations of motion for the gravitational field, sourced by minimally coupled components and non-minimally coupled dark matter, are
\begin{eqnarray}\label{eq|fieldeq}
        \frac{1}{8\pi\,G}\,G^{\mu\nu}& =& T_{\textsc{sm}}^{\mu\nu} + T_{\textsc{dm}}^{\mu\nu} - L^2\,\bigg[\left(G^{\mu\nu} + g^{\mu\nu}\,\Box - \nabla^{(\mu}\nabla^{\nu)}\right)\,T_{\textsc{dm}} - \Box\,T_{\textsc{dm}}^{\mu\nu} + 2\,\nabla^{\alpha}\,\nabla^{(\mu}\,T^{\nu)}_{\alpha}  + \nonumber\\
        &&- g^{\mu\nu}\nabla_{\alpha}\,\nabla_{\beta}\,T^{\alpha\beta}_{\textsc{dm}} - \frac{R}{2}\left(T_{\textsc{dm}}^{\mu\nu}-g^{\mu\nu}T_{\textsc{dm}}\right) - \frac{T_{\textsc{dm}}^{\mu\nu}}{T_{\textsc{dm}}}\,R_{\alpha\beta}\,T^{\alpha\beta}_{\textsc{dm}}\bigg]\,.
\end{eqnarray}
In addition, one has also the conservation equations for the minimally coupled fluids,
\begin{align}\label{eq|mincons}
    \nabla_\mu T^{\mu\nu}_{\textsc{sm}}=0\,,
\end{align}
since no direct interaction between dark matter and the other components is assumed.

Taking the covariant divergence of equation~\eqref{eq|fieldeq} and using the contracted Bianchi identities, $\nabla_\mu G^{\mu\nu}=0$ and equation~\eqref{eq|mincons}, one obtains a modified conservation equation for the dark matter component,
\begin{equation}\label{eq|Bianchi}
    \begin{split}
        \nabla_{\mu}T^{\mu\nu}_{\textsc{dm}} =&\ L^2\,\Bigg[2{R^{\nu}}_{\lambda}\nabla_{\mu}T^{\mu\lambda}_{\textsc{dm}}-\frac{1}{2}R\nabla^{\nu}\,T_{\textsc{dm}}+2T^{\mu\lambda}_{\textsc{dm}}\nabla_{\mu}{R^{\nu}}_{\lambda}- T^{\mu\lambda}_{\textsc{dm}}\nabla^{\nu}R_{\mu\lambda} +\\
        & -\nabla_{\mu}\bigg(\frac{R}{2}\left(T^{\mu\nu}_{\textsc{dm}}-g^{\mu\nu}T_{\textsc{dm}}\right) + \frac{T^{\mu\nu}_{\textsc{dm}}}{T_{\textsc{dm}}}\,R_{\alpha\beta}\,T^{\alpha\beta}_{\textsc{dm}}\bigg)\Bigg]\,,
    \end{split}
\end{equation}
expressing the simple fact that, due to the non-minimal coupling, the dark matter stress-energy tensor is not conserved independently.

\subsection{Background evolution}\label{subsec|background}

We adopt a Friedmann–Lema\^{i}tre–Robertson–Walker (FLRW) ansatz for the metric
\begin{equation}\label{eq|friedmannmetric}
ds^2 = -dt^2+a^2(t)\left(\frac{dr^2}{1-\kappa\,r^2}+r^2d\Omega^2\right),
\end{equation}
where $a(t)$ is the scale factor, normalized to unity at the present time $a_0:=a(t_0)=1$, and $\kappa=0,\pm R_0^{-2}$ is the Gaussian curvature of the hypersurface at $t=t_0$. We consider a $\Lambda$CDM cosmology in which the Universe is filled by four components: baryons $({}_{\textsc{b}})$, dark matter $({}_{\textsc{dm}})$, radiation composed of photons and neutrinos $({}_{\textsc{r}})$, and a cosmological constant $\Lambda$ $({}_\Lambda)$. All components are treated as perfect barotropic fluids with equations of state $p_n(t) = w_n\rho_n(t)$, where the subscript ${}_n$ labels the different components and $w_{\textsc{b}} = w_{\textsc{dm}} = 0, \, w_{\textsc{r}} = 1/3,\, \text{and } w_{\Lambda} = -1$. For simplicity, we also introduce the dimensionless variable
\begin{equation}\label{eq|defchi}
    \chi:=8\pi G L^2\rho_{\textsc{dm}}.
\end{equation}
This quantity measures the dark matter energy density in units of a critical scale $\rho_{\rm crit}=\left(8\pi G L^2\right)^{-1}$, so that $\chi\gg1$ corresponds to a regime in which the non-minimal coupling is relevant, while $\chi\ll1$ identifies the GR limit.

Substituting the ansatz~\eqref{eq|friedmannmetric} into the field equations~\eqref{eq|fieldeq}, we obtain two independent equations:
\begin{equation}\label{eq|friedmann}
    H^2 :=\left(\frac{\dot{a}}{a}\right)^2= \frac{8\pi G}{3}\frac{\rho_{\textsc{dm}} + \rho_{\textsc{b}} + \rho_{\textsc{r}} + \rho_{\Lambda}}{1+\chi} - \frac{\kappa}{a^2}\frac{1-\chi}{1+\chi}, 
\end{equation}
which is a modified Friedmann equation, and
\begin{equation}\label{eq|Raychaudhuri}
    \dot H+H^2:=\frac{\ddot{a}}{a}= -\frac{4\pi G}{3} \frac{\left(1-2\chi\right)\rho _{\textsc{dm}}+\left(1-2\chi\right)\rho_{\textsc{b}} + \left(2-\chi\right) \rho_{\textsc{r}}-\left(2+5\chi\right)\rho _{\Lambda}}{\left(1+\chi\right)^2}-\frac{\kappa}{a^2}\frac{3\chi}{\left(1+\chi\right)^2},
\end{equation}
which corresponds to the Raychaudhuri equation.
The time component of~\eqref{eq|Bianchi} yields the continuity equation
\begin{equation}\label{eq|conservation}
\left[1+\frac{3 \chi}{8\pi G \rho_{\textsc{dm}}}\left(H^2+\frac{\kappa}{a^2}\right)\right]\left(\dot\rho_{\textsc{dm}}+3H\rho_{\textsc{dm}}\right)+\dot\rho _{\textsc{b}}+3H\rho_{\textsc{b}} + \dot\rho _{\textsc{r}}+4H\rho_{\textsc{r}} + \dot\rho _{\Lambda }=0.
\end{equation}

As is standard in cosmology, we assume that there are no energy exchanges between the different components. Equation~\eqref{eq|conservation} then separates into four conditions, $\dot\rho_n + 3H(1+w_n)\rho_n=0$, one for each component, with solutions $\rho_n(t) = \rho_{n,0}\,a(t)^{-3(1+w_n)}$. For the minimally coupled fluids this behavior follows directly from the conservation laws~\eqref{eq|mincons}. For the dark matter component, however, the result is less obvious: despite the modified conservation equation~\eqref{eq|Bianchi}, the additional terms combine in such a way that the background evolution still reduces to the standard dust scaling.

To illustrate the cosmological evolution, and to keep track of the role played by the different components, it is useful to define a dynamical state parameter
\begin{equation}\label{effw}
    w := -\left(1+\frac{2}{3}\frac{\dot H}{H^2}\right).
\end{equation}
When $\dot{w}/H\ll 1$, i.e.~the state parameter changes slowly with respect to the Hubble time, the scale factor $a(t)$ evolves as
\begin{equation}
    a(t) \simeq \left(\frac{t}{t_0}\right)^{\frac{2}{3(1+w)}},
\end{equation}
and the evolution of $w$ thus has all the information on 
the time-dependence of the scale factor.

\begin{figure}[t!]
    \centering
    \includegraphics[width=\textwidth]{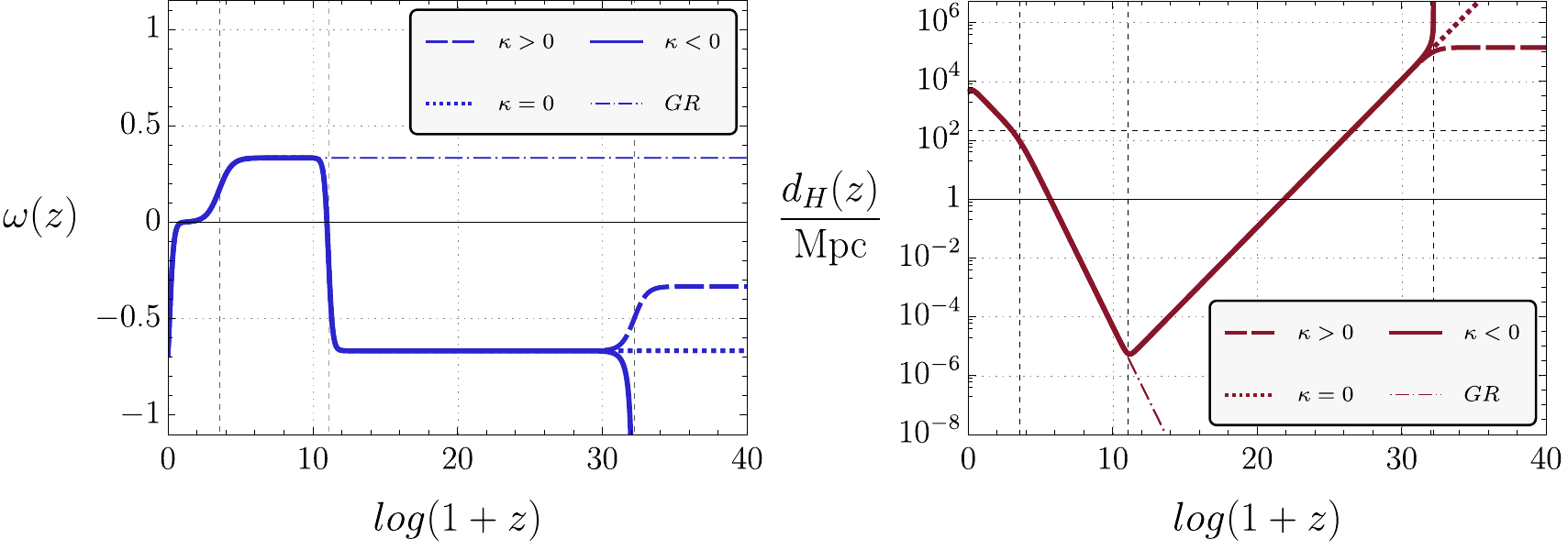}
    \caption{Left panel: the dynamical state parameter $w$ as a function of redshift; right panel: the comoving Hubble horizon as a function of redshift, as shown in~\cite{Silveravalle:2025yij}. The solid, dashed, and dotted lines correspond to NMC models with negative, positive, and zero spatial curvature, respectively, while the dot-dashed line shows the GR case. Vertical grey dotted lines indicate matter-radiation equality at redshift $\log\left(1+z_{\textsc{mre}}\right) \approx 3.54$, the onset of the NMC at redshift $\log\left(1+z_{\textsc{nmc}}\right)\approx 11.08$, and the beginning of the curvature era or the redshift of the bounce $\log\left(1+z_{\textsc{bnc}}\right) \approx 32.20$. The horizontal dashed line in the second panel marks the horizon at the epoch of recombination at redshift $\log\left(1+z_{\textsc{cmb}}\right) \approx 3.04$, which also intersects the horizon at $\log(1+z_{\textsc{end}})\approx 26.36$. 
    }
    \label{fig|w}
\end{figure}

The evolution of the dynamical state parameter as a function of redshift is shown in the left panel of figure~\ref{fig|w}. The solid, dashed, and dot-dashed lines correspond to NMC models with positive, zero, and negative curvature, respectively, while the dotted line shows the standard $\Lambda$CDM evolution. Up to a redshift $\log\left(1+z_{\textsc{nmc}}\right)\approx 11.08$, where $\chi\sim 1$, the evolution follows the GR behavior: there is a $\Lambda$-dominated epoch with $w\sim-1$ until $\log\left(1+z_{\Lambda}\right)\approx 0.11$, followed by a matter-dominated epoch with $w\sim 0$ up to matter-radiation equality at $\log\left(1+z_{\textsc{mre}}\right)\approx 3.54$, and a radiation-dominated epoch with $w\sim 1/3$ until the onset of the NMC. During the NMC-dominated epoch, radiation remains the dominant component, in the sense that $\rho_{\textsc{r}} \gg \rho_{\textsc{dm}}>\rho_{\textsc{b}} \gg \rho_\Lambda$, but the dynamical state parameter rapidly decreases from $w=1/3$ to $w=-2/3$, signaling a modified evolution due to the non-minimal coupling. In this phase, the scale factor evolves approximately as $a(t)\simeq t^2$, corresponding to an accelerated expansion of the Universe. If there is non-zero spatial curvature, at earlier times the Universe becomes curvature-dominated with $w\sim -1/3$ for $\kappa>0$, or undergoes a bounce with $H=0$ and $a\neq 0$ for $\kappa<0$. This occurs roughly when
\begin{equation}\label{eq|scalefacbounce}
    \frac{8\pi G}{3}\rho_{\textsc{r},{\textsc{bnc}}} \sim \frac{|\kappa|}{a_{\textsc{bnc}}^2}\chi_{\textsc{bnc}}\qquad\implies\qquad a_{\textsc{bnc}}\sim\frac{3|\kappa|\chi_0}{8\pi G \rho_{\textsc{r},0}},
\end{equation}
which corresponds to a redshift $\log\left(1+z_{\textsc{bnc}}\right) \gtrsim 32.20$ if we consider the constrain $|\kappa|< 1.2\times10^{-3}H_0^2$ set by the Planck collaboration.

The right panel of figure~\ref{fig|w} shows the evolution of the comoving Hubble horizon, defined as
\begin{equation}
    d_{\textsc{h}} := \frac{1}{aH},
\end{equation}
as a function of redshift. The horizontal dotted line indicates the value of the comoving Hubble horizon at recombination, which intersects the horizon again at $\log\left(1+z_{\textsc{end}}\right) \approx 26.54$. The accelerated expansion phase generated by the non-minimal coupling thus resolves both the horizon and flatness problems, provided that the cosmological parameters lie within the constraints of the Planck collaboration and that $L$ does not exceed our fiducial value.

\section{Theory of cosmological perturbations with a non-minimal coupling}\label{sec|theory}

A consistent cosmological model requires the analysis of perturbations. This analysis becomes even more important when the model is proposed as an alternative to inflation, since it must provide the primordial perturbations that evolve into CMB anisotropies and large-scale structure.
In this section, we present the equations governing perturbations in the presence of non-minimally coupled dark matter, and their solutions in the relevant regimes. A complete perturbation analysis is necessary, since the accelerated expansion phase of our model cannot be described within the slow-roll regime. The slow-roll parameters are
\begin{equation}
    \epsilon_{\textsc{h}}:=-\frac{\dot{H}}{H^2}=\frac{1}{2},\qquad\qquad\qquad\eta_{\textsc{h}}:=-\frac{1}{2}\frac{\ddot{H}}{\dot{H}H}=\frac{1}{2},
\end{equation}
which do not satisfy the condition $\epsilon_{\textsc{h}},\eta_{\textsc{h}}\ll1$.

In this and the following sections, we switch to conformal time $\mathrm{d}\eta=a^{-1}\mathrm{d} t$, so that
\begin{equation}\label{eq|conformal}
    \mathrm{d}s^2=a(\eta)^2\left(-\mathrm{d}\eta^2+\gamma_{ij}\mathrm{d}x^i\mathrm{d}x^j\right),\qquad f':=\frac{\mathrm{d}f}{\mathrm{d}\eta}=a\frac{\mathrm{d}f}{\mathrm{d}t}=a\dot{f},\qquad\mathcal{H}:=\frac{a'}{a}=\dot{a}=a H, 
\end{equation}
where $\mathcal{H}$ denotes the conformal Hubble parameter, and its inverse corresponds to the comoving Hubble radius $\mathcal{H}^{-1}=\left(aH\right)^{-1}=d_{\textsc{h}}$.

General perturbations of the metric and stress–energy tensor can be decomposed into scalar, vector, and tensor sectors, according to their transformation properties under spatial rotations. This decomposition is particularly useful at linear order, since the equations for the different sectors decouple, allowing their evolution to be studied independently.

Scalar perturbations represent fluctuations in the gravitational potential and in the energy density and pressure of the cosmic fluid. They are the most relevant sector for cosmological observations, as they generate the temperature anisotropies in the CMB and seed the formation of large-scale structure. The most general scalar perturbations of the metric and stress–energy tensor can be written as~\cite{Mukhanov:1990me}
\begin{equation}\label{eq|scalarpert}
    \delta g^{(\textsc{s})}_{\mu\nu}=a^2\begin{pmatrix}
        -2\phi & D_iB\\
        D_iB & -2\left(\psi\,\gamma_{ij}-D_iD_jE\right)
    \end{pmatrix},\quad {\delta T^\mu}_\nu^{(\textsc{s})}=\begin{pmatrix}
        -\delta\rho & \left(\rho+p\right)D_i\left(V-B\right)\\
        -\left(\rho+p\right)D^iV & \delta p\,{\delta^i}_j-D^iD_j\sigma
    \end{pmatrix},
\end{equation}
where $\phi$, $\psi$, $B$, $E$, $\delta\rho$, $\delta p$, $V$, and $\sigma$ are functions of the spacetime variables, and $D_i$ indicates the covariant spatial derivative. 

Vector perturbations induce rotational distortions of spacetime and frame-dragging effects. These modes are not observed at cosmological scales, and therefore a viable model must predict that they decay or remain negligible. The most general vector perturbations of the metric and stress–energy tensor can be written as
\begin{equation}\label{eq|vectorpert}
    \delta g^{(\textsc{v})}_{\mu\nu}=a^2\begin{pmatrix}
        0 & -S_i\\
        -S_i & D_jF_i+D_iF_j
    \end{pmatrix},\qquad\quad {\delta T^\mu}_\nu^{(\textsc{v})}=\begin{pmatrix}
        0 & \left(\rho+p\right)\left(U_i-S_i\right)\\
        -\left(\rho+p\right)U^i & D_j \nu^i+D^i \nu_j
    \end{pmatrix},
\end{equation}
where $S_i$, $F_i$, $U_i$ and $\nu_i$ are transverse three-vectors, satisfying $D_iS^i=D_iF^i=D_iU^i=D_i\nu^i=0$. 

Tensor perturbations induce transverse and traceless distortions of spacetime, i.e. primordial gravitational waves. They represent the propagating gravitational degrees of freedom in the absence of anisotropic stress, and their detection constrains models of the early Universe. The most general tensor perturbations of the metric and stress–energy tensor can be written as
\begin{equation}\label{eq|tensorpert}
    \delta g^{(\textsc{t})}_{\mu\nu}=a^2\begin{pmatrix}
        0 & 0\\
        0 & h_{ij}
    \end{pmatrix},\qquad\qquad  {\delta T^\mu}_\nu^{(\textsc{t})}=\begin{pmatrix}
        0 & 0\\
        0 & {\pi^i}_j
    \end{pmatrix},
\end{equation}
where $h_{ij}$ and $\pi_{ij}$ are symmetric, transverse, and traceless spatial tensors satisfying ${h^i}_i=D_ih^{ij}={\pi^i}_i=D_i\pi^{ij}=0$. 

In this work, we assume the absence of anisotropic stress, setting $\sigma=\nu_i=\pi_{ij}=0$, and restrict to adiabatic perturbations, so that $\delta p=c_s^2\delta\rho$.

\subsection{Scalar perturbations}\label{subsec|scalar}

Most cosmological phenomenology lies in scalar perturbations. Their dynamics can be complex, since the coupling between density fluctuations and the gravitational potential can produce feedback effects and lead to instabilities. They are also the only sector directly constrained by observations, and for this reason our analysis will mainly focus on them.

We first note that the functions defined in equations~\eqref{eq|scalarpert} are not gauge invariant. We introduce the gauge-invariant variables
\begin{equation}\label{eq|gaugeinv}
    \begin{split}
        \Phi:=&\ \phi+\mathcal{H}\left(B-E'\right)+\left(B-E'\right)',\qquad \Psi:=\psi-\mathcal{H}\left(B-E'\right),\\
        v_n:=&\ V_n+B-E',\qquad\qquad\qquad\qquad\ \ \ \delta_{n}:=\rho_n^{-1}\left(\delta\rho_n-3\left(1+{w_n}\right)\mathcal{H}\left(B-E'\right)\right),
    \end{split}
\end{equation}
where $n=\textsc{dm},\textsc{b},\textsc{r},\Lambda$. As in GR, using these gauge-invariant variables we obtain a system of equations equivalent to that in the longitudinal gauge, i.e.~with $B=E=0$. In particular, the $00$, $0i$, $ii$ and off-diagonal $ij$ components of the Einstein equations are
\begin{equation}\label{eq|perteinestein}
    \begin{split}
        -3\left(1+\chi\right)\mathcal{H}\left(\Psi'+\mathcal{H}\Phi\right)+\left(1-\chi\right)\left(\nabla^2\Psi+3\kappa\Psi\right)-\chi\left(2\mathcal{H}\nabla^2v_{\textsc{dm}}-\frac{3}{2}\left(\mathcal{H}^2-\kappa \right)\delta_{\textsc{dm}}\right) =& \\ 
        =4\pi Ga^2\displaystyle\sum_{n}\rho_n\delta_n, & \\[0.4cm]
        D_i\left(\left(1-\chi\right)\left(\Psi'+\mathcal{H}\Phi\right)-\chi\left(\frac{1}{2}\left(4\mathcal{H}'-\mathcal{H}^2+3\kappa\right)v_{\textsc{dm}}+\mathcal{H}\delta_{\textsc{dm}}\right)\right) =& \\
        =D_i\left(4\pi G a^2\displaystyle\sum_{n}(1+w_{n})\rho_n v_n\right), & \\[0.4cm]
        \left(1+\chi\right)\left(\Psi''+2\mathcal{H}\Psi'-\kappa\Psi+\mathcal{H}\Phi'+\left(2\mathcal{H}'+\mathcal{H}^2\right)\Phi-\frac{1}{2}\nabla^2\Psi+\frac{1}{2}\nabla^2\Phi\right)+&\\
         -\chi\left(3\mathcal{H}^2\Phi+\kappa\Psi+\nabla^2\Phi+\mathcal{H}\delta_{\textsc{dm}}'+\left(\mathcal{H}'-\mathcal{H}^2+\kappa\right)\delta_{\textsc{dm}}-\nabla^2v_{\textsc{dm}}'+\mathcal{H}\nabla^2v_{\textsc{dm}}'+\frac{1}{2}\nabla^2\delta_{\textsc{dm}}\right)=&\\
         =4\pi G a^2\displaystyle\sum_{n}w_{n}\rho_n\delta_n,& \\[0.4cm]
         D_iD_j\left(\frac{1}{2}\left(1-\chi\right)\left(\Psi-\Phi\right)-\chi \left(v_{\textsc{dm}}'-\mathcal{H}v_{\textsc{dm}}-\frac{1}{2} \delta_{\textsc{dm}}\right)\right) =0,&
    \end{split}
\end{equation}
where the sums are on the indices $n=\textsc{dm},\textsc{b},\textsc{r},\Lambda$, and we used that for barotropic fluids $c_{s,n}^2=w_n$. The conservation of the minimally coupled stress-energy tensors, namely those of radiation, baryons, and dark energy, leads to
\begin{equation}\label{eq|pertset}
    \begin{split}
        \delta_m'-\left(1+w_m\right)\left(3\Psi'+\nabla^2v_m\right)=0&\,, \\
        D_i\left(v_m'+\left(1-3w_m\right)\mathcal{H}v_m-\Phi-\frac{w_m}{1+w_m}\delta_m\right)=0&\,;
    \end{split}
\end{equation}
where $m=\textsc{b},\textsc{r},\Lambda$.  
Let us note that for the cosmological constant component ($w_\Lambda=-1$) the conservation of the stress-energy tensor implies $\delta_\Lambda=\delta_\Lambda'=0$. Instead of the conservation of the stress-energy tensor for dark matter, we must consider the divergence of the full system of equations of motion, which, using the Bianchi identities, implies
\begin{equation}\label{eq|pertbianchi}
    \begin{split}
        \delta_{\textsc{dm}}'-3\Psi'-\nabla^2v_{\textsc{dm}}=0&, \\[0.4cm]
        D_i\Bigg(v_{\textsc{dm}}'+\mathcal{H}v_{\textsc{dm}}-\Phi+\frac{\chi}{2\pi G a^2\rho_{\textsc{dm}}}\bigg(\Psi''+\frac{1}{2}\mathcal{H}\Psi'+\mathcal{H}\Phi'-\frac{1}{2}\nabla^2\Psi-\frac{3}{2}\kappa\Psi+&\\+\frac{1}{4}\left(4\mathcal{H}'-\mathcal{H}^2-3\kappa\right)\Phi
        +\frac{1}{4}\left(4\mathcal{H}'-\mathcal{H}^2-\kappa\right)v_{\textsc{dm}}'+\frac{1}{4}\left(4\mathcal{H}''-\left(6\mathcal{H}'-\mathcal{H}^2-\kappa\right)\mathcal{H}\right)v_{\textsc{dm}}\bigg) \Bigg)=0&.
    \end{split}
\end{equation}
Interestingly, the continuity equation for dark matter perturbations is left unchanged, while the Euler equation is significantly modified. As expected, in the absence of non-minimal coupling ($\chi\to0$), equations~(\ref{eq|perteinestein}--\ref{eq|pertbianchi}) reduce to those of GR.

For the rest of the paper, we will consider small spatial curvature $\kappa\sim 0$; this assumption is justified by the fact that the length scales of perturbations are much smaller than the curvature radius of the Universe, with the notable exception of the bounce, which, however, does not change the global evolution of perturbations, as we show in Appendix~\ref{app|bounce}. We also move to Fourier space $D_i\to i\, k_i,\ \nabla^2\to-k^2$ and consider the effective state parameter $w$ and sound speed $c_s^2$, defined as
\begin{equation}\label{eq|effstate}
    w:=-\left(1+\frac{2}{3}\frac{\dot{H}}{H^2}\right)=-\frac{1}{3}\left(1+2\frac{\mathcal{H}'}{\mathcal{H}^2}\right),\qquad\qquad c_s^2:=\frac{\displaystyle\sum_{n}c_{s,n}^2\rho_n\delta_n}{\displaystyle\sum_{n}\rho_n\delta_n}=\frac{\displaystyle\sum_{n}w_{n}\rho_n\delta_n}{\displaystyle\sum_{n}\rho_n\delta_n}.
\end{equation}
In particular, we approximate the evolution of the Universe as a sequence of epochs where $w$ and $c_s^2$ are constant (namely, a $\Lambda$-dominated epoch with $w\sim c_s^2\sim-1$, a matter-dominated epoch with $w\sim c_s^2\sim0$, a radiation-dominated epoch with $w\sim c_s^2\sim1/3$, and a NMC-dominated epoch with $w\sim-2/3$ and $c_s^2\sim 1/3$). With these definitions, all the information about the minimally coupled fluids is encoded in $w$ and $c_s^2$. 

After using the off-diagonal $ij$ component of equations~\eqref{eq|perteinestein} to solve for $\Phi$, the dynamics can be described using the two equations~\eqref{eq|pertbianchi} and the combination $\mathcal{E}_{ii}-c_s^2\mathcal{E}_{00}=0$ of equations~\eqref{eq|perteinestein}, which together form a system at first order in $\delta_{\textsc{dm}}$ and at second order in $\Psi$ and $v_{\textsc{dm}}$. The dynamics of baryon and radiation perturbations follow from equations~\eqref{eq|pertset}, once $\Phi$ and $\Psi$ have been determined by this system.

\subsubsection{Comoving curvature perturbation}\label{subsubsec|comoving}

It is standard in cosmology to track the evolution of metric perturbations using the comoving curvature perturbation, that is, the perturbation of the spatial metric $\psi$ (which in GR is equivalent to the Newtonian potential $\phi$ in the absence of anisotropic stress) as seen by an observer who is comoving with the cosmic fluid. Practically, it is defined as $\mathcal{R}:=\psi+\mathcal{H}\xi^0$, where $\xi^0$ is fixed by the comoving gauge, defined by $T_{0i}=0$:
\begin{equation}
    T_{0i}=\displaystyle\sum_n (1+w_n)\rho_n \left(v_n-\xi^0\right)=0\qquad\implies\qquad\xi^0=\frac{\sum_n (1+w_n)\rho_n v_n}{\sum_n (1+w_n)\rho_n},
\end{equation}
where we used the gauge transformation for the velocity fields $v_n\to v_n-\xi^0$. Using the $0i$ and the off-diagonal $ij$ components of equations~\eqref{eq|perteinestein}, and the background equations~\eqref{eq|friedmann}-~\eqref{eq|Raychaudhuri} expressed in conformal time, we obtain
\begin{equation}\label{eq|defcomo}
    \mathcal{R}=\Psi+\frac{2}{3}\frac{\Psi+\frac{\Psi'}{\mathcal{H}}-\chi\left(\Psi+\frac{\Psi'}{\mathcal{H}}-\left(\frac{7}{2}+3w\right)\mathcal{H}v_{\textsc{dm}}+2v_{\textsc{dm}}'\right)}{1+w+\chi\left(2+w\right)},
\end{equation}
which reduces to the standard definition in the GR limit $\chi\ll1$.

The final form of the equations is found by substituting $\Psi$ and $\Psi'$ with $\mathcal{R}$ and $\mathcal{R}'$. However, the full set of equations is very lengthy and not particularly instructive, so we do not report it here\footnote{It can be provided by the authors in the form of a \texttt{Mathematica} notebook on request.}. 
In the GR limit, we recover the standard equations
\begin{equation}\label{eq|eqcomogr}
    \begin{split}
        \mathcal{R}''+2\mathcal{H}\mathcal{R}'+c_s^2k^2\mathcal{R}=&\ 0,\\
        \delta_{\textsc{dm}}'+k^2 v_{\textsc{dm}}-\frac{9}{2}\left(1+c_s^2\right)\mathcal{H}\mathcal{R}-\frac{9}{4}\left(1+c_s^2\right)\left(5+3c_s^2\right)\mathcal{H}^2\frac{\mathcal{R}'}{c_s^2k^2}=&\ 0,\\
        v_{\textsc{dm}}'+\mathcal{H}v_{\textsc{dm}}+\frac{3}{2}\left(1+c_s^2\right)\mathcal{H}\frac{\mathcal{R}'}{c_s^2k^2}=&\ 0;
    \end{split}
\end{equation}
the equation for the comoving curvature perturbation can be integrated directly, yielding
\begin{equation}\label{eq|solcomogr}
    \mathcal{R}=C_{1}\frac{H_{-\nu}^{(1)}\left(c_s k\,\eta\right)}{\eta^{\nu}}+C_{2}\frac{H_{-\nu}^{(2)}\left(c_s k\,\eta\right)}{\eta^{\nu}},
\end{equation}
where $H_{\alpha}^{(1)}$ and $H_{\alpha}^{(2)}$ are Hankel functions of the first and second kind, $\nu=\frac{3}{2}\frac{1-w}{1+3w}$, and $C_{1}$ and $C_{2}$ are integration constants which depend on $w$ and $c_s^2$.

To investigate the effect of the non-minimal coupling on perturbations, we consider the NMC-dominated epoch by taking the limit $\chi\gg1$ and setting $w=-2/3$ and $c_s^2=1/3$. The equations of motion in this regime are
\begin{equation}\label{eq|eqcomonmc}
    \begin{split}
        \mathcal{R}''-2\mathcal{H}\frac{2k^2-3\mathcal{H}^2}{4k^2-3\mathcal{H}^2}\mathcal{R}'+\frac{4}{3}\frac{2k^2-3\mathcal{H}^2}{4k^2-3\mathcal{H}^2}\mathcal{R}-\frac{\mathcal{H}}{4}v_{\textsc{dm}}''-\frac{3}{4}\frac{\mathcal{H}^4}{4k^2-3\mathcal{H}^2}v_{\textsc{dm}}'-\frac{\mathcal{H}}{24}\left(4k^2-3\mathcal{H}^2\right)v_{\textsc{dm}}=&\ 0,\\
        \delta_{\textsc{dm}}'+3\mathcal{H}\frac{8k^2-3\mathcal{H}^2}{4k^2-3\mathcal{H}^2}v_{\textsc{dm}}'+\left(k^2+\frac{3\mathcal{H}^2}{2}\right)v_{\textsc{dm}}-\frac{36\mathcal{H}^2}{4k^2-3\mathcal{H}^2}\mathcal{R}'+\frac{24k^2\mathcal{H}}{4k^2-3\mathcal{H}^2}\mathcal{R}=&\ 0,\\
        v_{\textsc{dm}}'+\frac{5\mathcal{H}}{2}\frac{4k^2-3\mathcal{H}^2}{k^2+3\mathcal{H}^2}v_{\textsc{dm}}+\frac{1}{2}\frac{4k^2-3\mathcal{H}^2}{k^2+3\mathcal{H}^2}\delta_{\textsc{dm}}+\frac{2}{\mathcal{H}}\frac{2k^2-9\mathcal{H}^2}{k^2+3\mathcal{H}^2}\mathcal{R}'+\frac{3\left(2k^2+\mathcal{H}^2\right)}{k^2+3\mathcal{H}^2}\mathcal{R}=&\ 0.
    \end{split}
\end{equation}

While this system still appears quite involved, it can be solved analytically after introducing an appropriate variable, as shown in Appendix~\ref{app|nmccomoving}. The comoving curvature perturbation in the NMC-dominated epoch then takes the form
\begin{equation}\label{eq|solcomonmc}
    \mathcal{R}=C_{1,\textsc{nmc}}k^2\frac{\mathrm{e}^{i\sqrt{\frac{2}{3}}k\eta}}{\eta}+C_{2,\textsc{nmc}}k^2\frac{\mathrm{e}^{-i\sqrt{\frac{2}{3}}k\eta}}{\eta}+C_{3,\textsc{nmc}}\frac{\mathrm{e}^{k\eta}}{\eta}+C_{4,\textsc{nmc}}\frac{\mathrm{e}^{-k\eta}}{\eta},
\end{equation}
where the $C_{i,\textsc{nmc}}$ are the integration constants of the NMC epoch. The real exponential factors are due to non-oscillatory effects in the velocity field, and their associated integration constants can in principle be set to zero; nonetheless, during the NMC-dominated epoch the conformal time $\eta$ decreases toward $\eta\ll1/k$, and therefore they contribute to the comoving curvature perturbation as a constant term. We also note that the solution in equation~\eqref{eq|solcomonmc} has the familiar form of oscillations modulated by the conformal time, but with an additional factor $k^2$, which modifies the scale dependence of the perturbations.

To understand the physical behavior of perturbations, it is convenient to distinguish between a super-horizon regime, in which the perturbation length scales are larger than the Hubble radius $\left(k\ll\mathcal{H}\right)$, and a deep sub-horizon regime, in which the perturbations are much smaller than the horizon and can therefore interact dynamically $\left(k\gg\mathcal{H}\right)$.
In the super-horizon regime, the solutions in equations~\eqref{eq|solcomogr} and~\eqref{eq|solcomonmc} reduce respectively to
\begin{equation}\label{eq|comosuphgr}
        \mathcal{R}_{k\ll\mathcal{H}}\sim \frac{i\Gamma(-\nu)\left(\frac{1}{2}c_s k\right)^\nu \left(C_{2}-C_{1}\right)}{\pi}+\frac{C_{1}+C_{2}+i\cot(\pi\nu)\left(C_{2}-C_{1}\right)}{\Gamma(1-\nu)\left(\frac{1}{2}c_s k\right)^{\nu}}\eta^{-2\nu},
\end{equation}
where $\Gamma$ is Euler gamma function, and
\begin{equation}\label{eq|comosuphnmc}
\begin{split}
         \mathcal{R}_{k\ll\mathcal{H}}\sim &\ k\left(i\sqrt{\frac{2}{3}}k^2\left(C_{1,\textsc{nmc}}-C_{2,\textsc{nmc}}\right)+C_{3,\textsc{nmc}}-C_{4,\textsc{nmc}}\right)+\\
         &+\left(k^2\left(C_{1,\textsc{nmc}}+C_{2,\textsc{nmc}}\right)+C_{3,\textsc{nmc}}+C_{4,\textsc{nmc}}\right)\eta^{-1}.   
\end{split}
\end{equation}
In both cases, the solution consists of a constant mode and a time-dependent mode. However, while in the GR case the time-dependent mode decays in an expanding Universe as $\eta^{-2\nu}\propto a^{-\frac{3}{2}\left(1-w\right)}$, in the NMC case it instead grows as $\eta^{-1}\propto a^{1/2}$. In the deep sub-horizon regime, equations~\eqref{eq|solcomogr} and~\eqref{eq|solcomonmc} reduce respectively to
\begin{equation}\label{eq|comosubhgr}
        \mathcal{R}_{k\gg\mathcal{H}}\sim C_{1}\frac{\mathrm{e}^{-i\frac{\pi}{4}(1-2\nu)}}{\sqrt{\frac{\pi}{2}c_s\,k}}\frac{\mathrm{e}^{ic_s k \eta}}{\eta^{\frac{1}{2}+\nu}} +C_{2}\frac{\mathrm{e}^{i\frac{\pi}{4}(1-2\nu)}}{\sqrt{\frac{\pi}{2}c_s\,k}}\frac{\mathrm{e}^{-ic_s k \eta}}{\eta^{\frac{1}{2}+\nu}},
\end{equation}
and
\begin{equation}\label{eq|comosubhnmc}
        \mathcal{R}_{k\gg\mathcal{H}}\sim C_{1,\textsc{nmc}}k^2\frac{\mathrm{e}^{i\sqrt{\frac{2}{3}}k\eta}}{\eta}+C_{2,\textsc{nmc}}k^2\frac{\mathrm{e}^{-i\sqrt{\frac{2}{3}}k\eta}}{\eta}+C_{(3,4),\textsc{nmc}}\frac{\mathrm{e}^{\pm k\eta}}{\eta},
\end{equation}
where the last term depends on whether $\eta$ is positive or negative. To preserve the validity of the linear approximation in the sub-horizon regime, we set $C_{3,\textsc{nmc}}=0$ in an expanding Universe and $C_{4,\textsc{nmc}}=0$ in an eventual contracting phase. In both cases, the solutions have oscillations modulated by a time-dependent amplitude. As in the super-horizon regime, in the GR case the amplitude decays in an expanding Universe as $\eta^{-\frac{1+2\nu}{2}}\propto a^{-1}$, while in the NMC case it instead grows as $\eta^{-1}\propto a^{1/2}$.

\subsection{Vector perturbations}\label{subsec|vecotr}

For vector perturbations, gauge-invariant variables are defined as
\begin{equation}\label{eq|gaugeinvV}
        \Psi_i:= S_i+F_i',\qquad\qquad V_{n,i}:=U_{n,i}-F_i',
\end{equation}
and the equations are equivalent to those obtained in the gauge $F_i=0$. The non-vanishing components of the Einstein equations, namely the $0i$ and $ij$ ones, are
\begin{equation}\label{eq|pertVeinstein}
    \begin{split}
        \left[\left(1+2\chi\right)\mathcal{H}^2-\left(1-\chi\right)\mathcal{H}'\right]\Psi_i-\frac{1}{2}\chi\left(4\mathcal{H}'-\mathcal{H}^2\right)V_{\textsc{dm},i}-\frac{1}{4}\left(1-3\chi\right)\nabla^2\Psi_i-\frac{1}{2}\chi\nabla^2V_{\textsc{dm},i}=&\\
        =4\pi G a^2\displaystyle\sum_{n}(1+w_{n})\rho_n V_{n,i},&\\[0.4cm]
        \left(1-\chi\right)\left(D_i\Psi_j'+D_j\Psi_i'\right)+\left(2+\chi\right)\mathcal{H}\left(D_i\Psi_j+D_j\Psi_i\right)+2\chi\left(D_iV_{\textsc{dm},j}'+D_jV_{\textsc{dm},i}'\right)+&\\
        -2\chi\mathcal{H}\left(D_iV_{\textsc{dm},j}+D_jV_{\textsc{dm},i}\right)=0.&
    \end{split}
\end{equation}
The $ii$ components take a particularly simple form:
\begin{equation}\label{eq|pertVeinsteinii}
    \begin{split}
        D_i\left[\left(1-\chi\right)\Psi_i'+\left(2+\chi\right)\mathcal{H}\Psi_i+2\chi V_{\textsc{dm},i}'-2\chi\mathcal{H}V_{\textsc{dm},i}\right]=0.
    \end{split}
\end{equation}
The conservation of the minimally coupled stress-energy tensors reads
\begin{equation}\label{eq|pertVset}
    \begin{split}
        \left(1+w_m\right)\left[V_{m,i}'-\Psi_i'+\left(1-3w_m\right)\mathcal{H}\left(V_{m,i}-\Psi_i\right)\right]=0,
    \end{split}
\end{equation}
where $m=\textsc{b},\textsc{r},\Lambda$, which vanishes identically for 
a cosmological constant ($w_\Lambda=-1$), since it does not support vector perturbations (see equation~\eqref{eq|vectorpert}). Taking the divergence of the equations of motion and using the contracted Bianchi identities, one obtains a modified conservation equation for dark matter:
\begin{equation}\label{eq|pertVbianchi}
    \begin{split}
        V_{\textsc{dm},i}'-\Psi_i'+\mathcal{H}\left(V_{\textsc{dm},i}-\Psi_i\right)-\frac{\chi}{8\pi G a^2\rho_{\textsc{dm}}}\bigg[\left(\mathcal{H}^2-4\mathcal{H}'\right)\left(V_{\textsc{dm},i}'-\Psi_i'\right)+&\\
        -\left(4\mathcal{H}''-6\mathcal{H}'\mathcal{H}+\mathcal{H}^3\right)\left(V_{\textsc{dm},i}-\Psi_i\right)+\nabla^2\Psi_i'-\mathcal{H}\nabla^2\Psi_i\bigg]=0,&
    \end{split}
\end{equation}
The conservation equations~\eqref{eq|pertVset} can be readily integrated to give
\begin{equation}\label{eq|solVfluid}
    V_{m,i}=\frac{C_{m,i}}{a^{1-3w}}+\Psi_i,
\end{equation}
and the modified conservation equation for dark matter can be integrated in Fourier space as
\begin{equation}\label{eq|solVdm}
    V_{\textsc{dm},i}=\frac{C_{\textsc{dm},i}}{a}\frac{1}{1-\frac{\chi}{8\pi G a^2\rho_{\textsc{dm}}}\left(\mathcal{H}^2-4\mathcal{H}'\right)}+\left(1-k^2\frac{\frac{\chi}{8\pi G a^2\rho_{\textsc{dm}}}}{1-\frac{\chi}{8\pi G a^2\rho_{\textsc{dm}}}\left(\mathcal{H}^2-4\mathcal{H}'\right)}\right)\Psi_i.
\end{equation}
Finally, using equation~\eqref{eq|solVdm}, the $ii$ components of the Einstein equations~\eqref{eq|pertVeinsteinii} can be integrated to obtain
\begin{equation}\label{eq|solVg}
    \Psi_i=-\frac{C_{\textsc{dm},i}}{a}\frac{\frac{2\chi}{1+\chi}}{1-\frac{\chi}{8\pi G a^2\rho_{\textsc{dm}}}\left(\mathcal{H}^2-4\mathcal{H}'+\frac{2\chi}{1+\chi}k^2\right)}+\frac{C_{\scriptstyle{\Psi},i}}{\left(1+\chi\right)a^2}\frac{1-\frac{\chi}{8\pi G a^2\rho_{\textsc{dm}}}\left(\mathcal{H}^2-4\mathcal{H}'\right)}{1-\frac{\chi}{8\pi G a^2\rho_{\textsc{dm}}}\left(\mathcal{H}^2-4\mathcal{H}'+\frac{2\chi}{1+\chi}k^2\right)},
\end{equation}
while the $0i$ components impose the constraints on the integration constants
\begin{equation}\label{eq|solVconstrint}
    C_{\scriptstyle{\Psi},i}=\frac{16\pi G}{k^2}\displaystyle\sum_n \left(1+w_n\right)\rho_{n,0}C_{n,i}.
\end{equation}

These perturbations have a singular behavior when the denominator in equation~\eqref{eq|solVg} vanishes. This occurs at a scale factor
\begin{equation}\label{eq|aVsing}
    a_s\sim\frac{3\chi_0k^2}{4\pi G\rho_{\textsc{r},0}}=6\frac{\rho_{\textsc{dm},0}}{\rho_{\textsc{r},0}}L^2k^2.
\end{equation}
While this singularity does not affect the global evolution of the perturbations, we restrict the analysis to modes generated after this point.

Even if a full analytical form is present, to study the evolution of vector perturbations it is instructive to consider separately the GR limit and the NMC-dominated phase. The perturbations of the minimally coupled fluids, described by equation~\eqref{eq|solVfluid}, follow the same evolution as the gravitational perturbation $\Psi_i$, up to the addition of a constant or decaying mode; for this reason, we focus only on the evolution of the gravitational and dark matter perturbations. In the GR limit, they have the standard behavior
\begin{equation}\label{eq|evoVgr}
    \Psi_i\sim\frac{C_{\scriptstyle{\Psi},i}}{a^2},\qquad\qquad V_{\textsc{dm},i}\sim\frac{C_{\textsc{dm},i}}{a}+\frac{C_{\scriptstyle{\Psi},i}}{a^2}.
\end{equation}
In the NMC-dominated epoch, they instead evolve as
\begin{eqnarray}\label{eq|evoVnmc}
    \Psi_i&\sim&\frac{8\pi G \rho_{\textsc{dm},0}C_{\textsc{dm},i}}{\chi_0 k^2}a=\frac{C_{\textsc{dm},i}}{L^2k^2}a\,,\nonumber \\V_{\textsc{dm},i}&\sim&\frac{1}{2}\left(\frac{C_{\scriptstyle{\Psi},i}}{\chi_0}+\frac{8\pi G \rho_{\textsc{dm},0}C_{\textsc{dm},i}}{\chi_0 k^2}\right)a=\frac{1}{2}\left(\frac{C_{\scriptstyle{\Psi},i}}{\chi_0}+\frac{C_{\textsc{dm},i}}{L^2k^2}\right)a\,.
\end{eqnarray}
The key difference with respect to the standard case is therefore a growth proportional to the scale factor in the early Universe. 

If we consider gravitational perturbations generated after the singular point in equation~\eqref{eq|aVsing}, their amplitude at recombination scales as
\begin{equation}\label{eq|evoVg}
    \Psi_i(a_{\textsc{cmb}})\propto \Psi_i(a_s)\frac{a_{\textsc{nmc}}}{a_s}\frac{a_{\textsc{nmc}}^2}{a_{\textsc{cmb}}^2}\sim\Psi_i(a_s)\frac{4\pi G\rho_{\textsc{r},0}}{3}\frac{1}{k^2}\frac{1}{a_{\textsc{cmb}}^2},
\end{equation}
where we used $a_{\textsc{nmc}}=\left(8\pi G L^2\rho_{\textsc{dm},0}\right)^{1/3}$ for which $\left.\chi\right|_{a=a_{\textsc{nmc}}}=1$. Perturbations are enhanced with respect to their initial value for wave numbers
\begin{equation}\label{eq|maxVk}
    k<\frac{1}{a_{\textsc{cmb}}}\sqrt{\frac{4\pi G\rho_{\textsc{r},0}}{3}}=(1+z_{\textsc{cmb}})\sqrt{\frac{\Omega_{\textsc{r},0}}{2}}H_0\sim1.65\times10^{-3}\mathrm{Mpc}^{-1}.
\end{equation}
If the initial perturbations are small (e.g. of quantum origin), all modes that are observationally relevant today remain of negligible amplitude.

\subsection{Tensor perturbations}\label{subsec|tensor}

To derive the equations for tensor perturbations, it is convenient to express them as
\begin{equation}\label{eq|Tmetric}
    h_{ij} = \begin{pmatrix}
        h_+ & h_\times & 0 \\
        h_\times & -h_+ & 0 \\
        0 & 0 & 0
    \end{pmatrix}.
\end{equation}
The field equations take a particularly simple form by combining the $(1,1)$ and $(2,2)$ components. In this way, we obtain
\begin{equation}\label{eq|h_+}
    (1+\chi)h_+'' + (2-\chi)\mathcal{H}\,h_+' + (1-\chi)\,k^2\,h_+ = 0\,,
\end{equation}
and an analogous equation for $h_{\times}$. In the GR limit, these equations reduce to
\begin{equation}\label{eq|GRT}
    h_{+/\times}'' + 2\,\mathcal{H}\,h_{+/\times}' + k^2\,h_{+/\times} = 0\,.
\end{equation}
Thus, the two polarization modes $h_+$ and $h_\times$ satisfy the same equation as the comoving curvature perturbation $\mathcal{R}$ in eq.~\eqref{eq|eqcomogr}, the only difference being that tensor perturbations travel at the speed of light $c_s^2=1$. The solution is therefore
\begin{equation}\label{eq|solTgr}
    h_{+/\times}=C_{1}^{+/\times}\frac{H_{-\nu}^{(1)}\left(k\,\eta\right)}{\eta^{\nu}}+C_{2}^{+/\times}\frac{H_{-\nu}^{(2)}\left(k\,\eta\right)}{\eta^{\nu}},
\end{equation}
where $H_{\alpha}^{(1)}$ and $H_{\alpha}^{(2)}$ are Bessel functions of the first and second kind, $\nu=\frac{3}{2}\frac{1-w}{1+3w}$, and $C_{1}^{+/\times}$ and $C_{2}^{+/\times}$ are integration constants which depend on $w$.

On the other hand, in the NMC limit tensor perturbations satisfy
\begin{equation}\label{eq|nmcT}
    h_{+/\times}'' -\mathcal{H}\,h_{+/\times}' - k^2\,h_{+/\times} = 0\,,
\end{equation}
which can be solved immediately, once we recall that in the NMC-dominated epoch $\mathcal{H}=-2/\eta$, to give
\begin{equation}\label{eq|solTnmc}
    h_{+/\times}=C_{1,\textsc{nmc}}^{+/\times}\frac{\mathrm{e}^{k\eta}}{\eta}+C_{2,\textsc{nmc}}^{+/\times}\frac{\mathrm{e}^{-k\eta}}{\eta},
\end{equation}
where $C_{1,\textsc{nmc}}^{+/\times}$ and $C_{2,\textsc{nmc}}^{+/\times}$ are the integration constants during the NMC epoch. The absence of oscillatory sub-horizon modes indicates that a standard adiabatic vacuum cannot be defined during the NMC-dominated epoch; consequently, tensor perturbations cannot be generated from vacuum quantum fluctuations, but can only correspond to the classical evolution of pre-existing perturbations.

In the super-horizon regime, the solutions in equations~\eqref{eq|solTgr} and~\eqref{eq|solTnmc} reduce respectively to
\begin{equation}\label{eq|Tsuphgr}
        h_{+/\times}^{k\ll\mathcal{H}}\sim \frac{i\Gamma(-\nu)\left(\frac{1}{2} k\right)^\nu \left(C_{2}^{+/\times}-C_{1}^{+/\times}\right)}{\pi}+\frac{C_{1}^{+/\times}+C_{2}^{+/\times}+i\cot(\pi\nu)\left(C_{2}^{+/\times}-C_{1}^{+/\times}\right)}{\Gamma(1-\nu)\left(\frac{1}{2} k\right)^{\nu}}\eta^{-2\nu},
\end{equation}
and
\begin{equation}\label{eq|Tsuphnmc}
        h_{+/\times}^{k\ll\mathcal{H}}\sim \frac{1}{2}(C^{+/\times}_{1,\textsc{nmc}}-C^{+/\times}_{2,\textsc{nmc}})+\left(C^{+/\times}_{1,\textsc{nmc}}+C^{+/\times}_{2,\textsc{nmc}}\right)\eta^{-1},
\end{equation}
which evolve precisely as scalar perturbations with speed $c_s^2=1$. In the deep sub-horizon regime, equations~\eqref{eq|solTgr} and~\eqref{eq|solTnmc} reduce to
\begin{equation}\label{eq|Tsubh}
        h_{+/\times}^{k\gg\mathcal{H}}\sim\ C^{+/\times}_{1}\frac{\mathrm{e}^{-i\frac{\pi}{4}(1+2\nu)}}{\sqrt{\frac{\pi}{2}\,k}}\frac{\mathrm{e}^{i k \eta}}{\eta^{\frac{1}{2}-\nu}} +C^{+/\times}_{2}\frac{\mathrm{e}^{i\frac{\pi}{4}(1+2\nu)}}{\sqrt{\frac{\pi}{2}\,k}}\frac{\mathrm{e}^{-i k \eta}}{\eta^{\frac{1}{2}-\nu}},
\end{equation}
and
\begin{equation}\label{eq|Tsubhnmc}
        h_{+/\times}^{k\gg\mathcal{H}}\sim C^{+/\times}_{(1,2),\textsc{nmc}}\frac{\mathrm{e}^{\pm k\eta}}{\eta}.
\end{equation}
Also in this case they evolve as scalar perturbations, with the notable difference that requiring the validity of the linear approximation sets $h_{+/\times}^{k\gg\mathcal{H}}\sim 0$ in the NMC-dominated epoch, and no primordial gravitational waves are generated in the early Universe.

\section{Generation of perturbations and primordial power spectrum}\label{sec|evolution}

The solutions derived in the previous section provide the necessary ingredients to study the generation of primordial perturbations. These perturbations set the initial conditions for cosmological anisotropies and can therefore be constrained through observations of the CMB and large-scale structure.

In the standard inflationary scenario, primordial perturbations originate from quantum fluctuations generated in the early Universe. During the accelerated expansion phase, the comoving horizon shrinks, so that fluctuations initially inside the horizon are stretched to super-horizon scales, where they effectively freeze. After the accelerated phase ends, they re-enter the horizon and source the observed anisotropies.

At the practical level, introducing the Mukhanov variable $u=a\mathcal{R}$~\cite{Mukhanov:1988jd}, the action at second order in perturbation theory reduces to that of a scalar field in flat spacetime with a time-dependent effective mass. The quantization of this scalar field fixes the normalization of the Fourier modes once an initial state is specified, usually taken to be the Bunch--Davies vacuum~\cite{Bunch:1978yq}. On super-horizon scales the comoving curvature perturbation $\mathcal{R}$ remains constant, implying $u\propto a$. Modes with larger wavelengths exit the horizon earlier and therefore undergo a longer amplification. Since horizon crossing is defined by $k\sim\mathcal{H}$, this translates into a characteristic dependence of the perturbation amplitude on the wavenumber $k$.

In the next subsections we adapt this procedure to our model, first introducing variables suitable for quantization and then following their evolution up to the epoch probed by the CMB. In the first subsection we consider the case in which primordial fluctuations are generated during the accelerated expansion phase. Once suitable variables are identified, the calculation is relatively straightforward, but the resulting predictions are incompatible with observations. In the second subsection we consider the case in which primordial fluctuations are generated during a contracting phase in a bouncing scenario. The evolution is considerably more involved, but we find that it is possible to obtain a power spectrum compatible with current observations. This requires an excited initial state rather than the Bunch--Davies vacuum. While such a choice does not appear unreasonable in a matter or radiation dominated contracting Universe, it substantially complicates the construction of the primordial spectrum.

The quantity directly related to observations is the power spectrum of perturbations. The power spectrum $\mathcal{P}_p(k)$ of a quantity $p(\mathbf{x},t)$ is defined as
\begin{equation}\label{eq|defps}
    \int_0^\infty\frac{\mathrm{d}k}{k}\mathcal{P}_p(k):=\langle0|p(\mathbf{x},t)p(\mathbf{x}+\Delta\mathbf{x},t)|0\rangle=\int_0^\infty\frac{\mathrm{d}^3k}{(2\pi)^3}|p_\mathbf{k}|^2 \quad \implies\quad \mathcal{P}_p(k):=\frac{k^3}{(2\pi)^2}|p_\mathbf{k}|^2,
\end{equation}
where the $p_\mathbf{k}$ are the Fourier modes with wavenumber $\mathbf{k}$. The main cosmological observables are the scalar spectral index
\begin{equation}\label{eq|defns}
    n_{\rm s}:=1+\frac{\mathrm{d}\ln\mathcal{P}_\mathcal{R}(k)}{\mathrm{d}\ln k}\,,
\end{equation}
where $\mathcal{P}_\mathcal{R}(k)$ is the power spectrum of comoving curvature perturbations, and the tensor-to-scalar ratio
\begin{equation}\label{eq|defr}
    r :=\frac{\mathcal{P}_h}{\mathcal{P}_\mathcal{R}}\,,
\end{equation}
where $\mathcal{P}_h$ is the power spectrum of tensor perturbations $h_{+/\times}$. The constraints from Planck collaboration are $n_{\rm s}= 0.9665\pm 0.0038$ and $r<0.036$, together with negligible vector perturbations~\cite{Planck}.

\subsection{Power spectrum within inflationary scenario}\label{subsec|inflation}

We first consider the case in which primordial fluctuations are generated during the accelerated expansion phase, so that the NMC-dominated epoch effectively plays the role of an inflationary era. 
In standard cosmological perturbation theory, the Mukhanov variable is derived from the action at second order in perturbations, which identifies the canonical degree of freedom and its normalization. While this would be the correct procedure to establish the canonical structure of the theory, deriving the full quadratic action for the present non-minimally coupled fluid system is technically involved and lies beyond the scope of this work. We therefore adopt a more pragmatic approach and work directly with the equations of motion, introducing a field redefinition that recasts the perturbation equation into a Mukhanov–Sasaki form.

Defining the variable\
\begin{equation}\label{eq|mukvar}
    u=\frac{1}{a}\left|1-\frac{4}{3}\frac{k^2}{\mathcal{H}^2}\right|^{-1/2}\left(\mathcal{R}-\frac{\mathcal{H}}{4}v_{\textsc{dm}}\right),
\end{equation}
equation~\eqref{eq|eqcomonmc} reduces to
\begin{equation}\label{eq|mukeq}
    u''+\left\{\frac{2}{3}k^2-\left[\frac{1}{2}+\left(\frac{\frac{k}{\mathcal{H}}}{\frac{4}{3}\frac{k^2}{\mathcal{H}^2}-1}\right)^2\right]\mathcal{H}^2\right\}u=0,
\end{equation}
which has the form of a Klein--Gordon equation for modes of wavenumber $\tilde{k}=\sqrt{\frac{2}{3}}k$ with a time-dependent effective mass. The field redefinition above reproduces the correct dynamical equation for a scalar degree of freedom, but it does not by itself fix its canonical normalization. In fact, any variable $\tilde{u}=C_k u$, with $C_k$ independent of $\eta$, satisfies the same equation~\eqref{eq|mukeq}, while having a different normalization. In particular, if the canonically normalized field satisfies $u\bar{u}'-u'\bar{u}=2i$, implying $u\propto k^{-1/2}$ in the sub-horizon regime, all the rescaled variables satisfy $\tilde{u}\bar{\tilde{u}}'-\tilde{u}'\bar{\tilde{u}}=2iC_k^2$. Therefore, if $C_k$ depends on the wavenumber, the $k^{-1/2}$ dependence of the canonical variable is not preserved.

To estimate the normalization of the quantized variable in the NMC-dominated regime, we start by defining the variable
\begin{equation}\label{eq|mukvargen}
    u_\zeta=\zeta\left(\mathcal{R}-\frac{\left(1+\chi\right)\chi\left(1+2w\right)\mathcal{H}}{\left(1-\chi\right)\left(1+w+\left(2+w\right)\chi\right)}v_{\textsc{dm}}\right),
\end{equation}
where $\zeta$ is chosen to remove the first-order derivative term from the equation of motion. This variable satisfies an equation which can formally be written as
\begin{equation}
  u_\zeta''+f_1\left(\eta,k\right)u_\zeta=f_2\left(\eta,k\right)v_{\textsc{dm}}+f_3\left(\eta,k\right)\delta_{\textsc{dm}},
\end{equation}
which is a Klein--Gordon-like equation with a source, and reduces to the standard Mukhanov--Sasaki equation in the GR limit and to equation~\eqref{eq|mukeq} in the NMC limit.

We have integrated numerically the function $\zeta$, imposing that the variable $u_\zeta$ reduces to the standard Mukhanov variable in the GR limit, that is, $\zeta\to a$ as $\chi\to0$. Since the Mukhanov variable is canonically normalized in the GR regime, this fixes the normalization in that limit. We then use the resulting evolution of $\zeta$ to infer the corresponding normalization in the NMC-dominated epoch. This does not constitute an independent derivation of the canonical normalization in the NMC regime, but provides a prescription for extending the known GR normalization into that regime. Interestingly, this also provides an indication that no ghost instability is encountered, since the normalization function satisfies a differential equation of the form $\zeta'(\eta)=f(\eta)\zeta(\eta)$ and therefore preserves the positive sign fixed by its GR-limit normalization.

\begin{figure}[t!]
    \centering
    \includegraphics[width=0.92\textwidth]{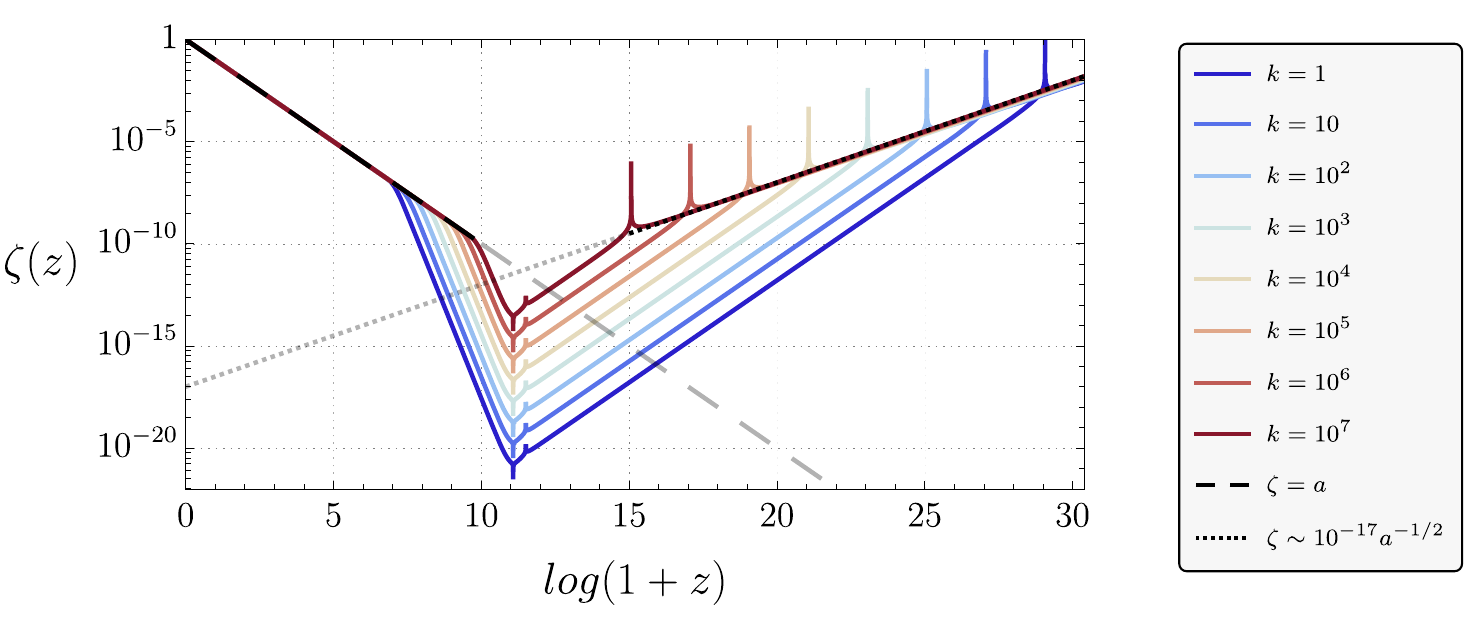}
    \caption{Transformation function between the comoving curvature perturbation $\mathcal{R}$ and Mukhanov variable $u$, as a function of redshift. The transformation depends on the wavenumber $k$ in the super-horizon regime, but approaches the standard $\zeta=a$ in the GR limit and $\zeta\propto a^{-1/2}$ in the sub-horizon NMC-dominated regime.}
    \label{fig|normalization}
\end{figure}

Figure~\ref{fig|normalization} shows the dependence of the transformation function $\zeta$ on the redshift and the wavenumber $k$. The asymptotic behaviour of the numerically determined $\zeta$ can then be compared with the NMC expression in equation\eqref{eq|mukvar}. By fitting the final part of $\zeta$ to this form with a normalization factor $C_k$, we infer that the variable with the normalization prescribed above behaves in the NMC limit as
\begin{equation}\label{eq|mukvarcn}
    u_{\textsc{cn}}\sim 10^{-32}\frac{k}{a}\left|1-\frac{4}{3}\frac{k^2}{\mathcal{H}^2}\right|^{-1/2}\left(\mathcal{R}-\frac{\mathcal{H}}{4}v_{\textsc{dm}}\right),
\end{equation}
corresponding to a $C_k\propto k$ factor that is absent in the na\"ive definition in equation~\eqref{eq|mukvar}.

\begin{figure}[t!]
    \centering
    \includegraphics[width=0.8\textwidth]{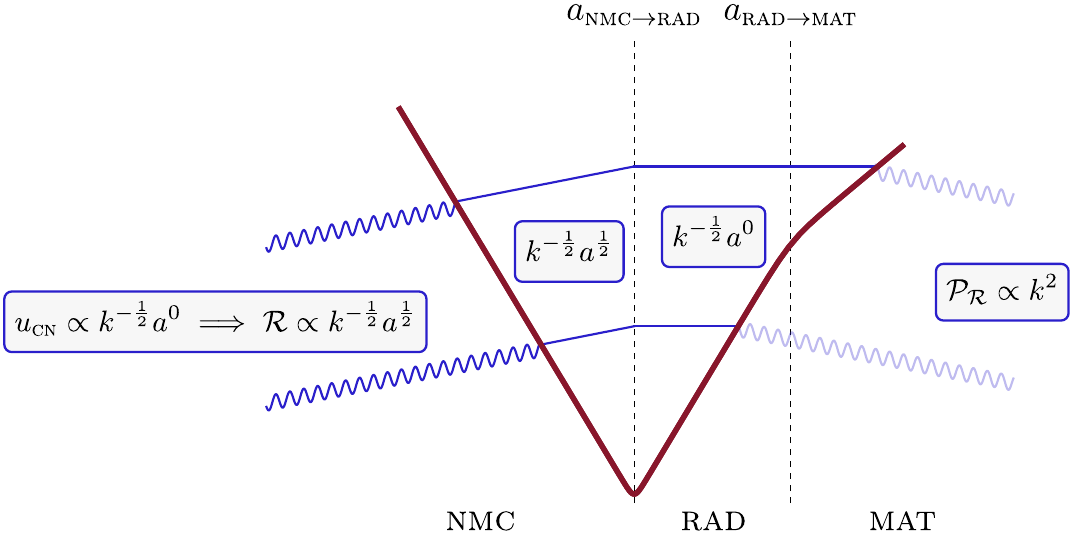}
    \caption{Pictorial representation of the evolution of perturbations generated during an inflationary era, showing their dependence on the wavenumber $k$ and on the scale factor $a$. All modes originate from a quantum state normalized as $k^{-1/2}$, which implies a $k^{-1/2}$-dependence also for the comoving curvature perturbation. The right-hand side of the figure shows the resulting $k$-dependence of the power spectrum.}
    \label{fig|inflation}
\end{figure}

Fluctuations are in the quantum regime when they are deeply sub-horizon, where the equation simplifies and admits the solution

\begin{equation}\label{eq|muksol}
    u_{\textsc{cn}}''+\frac{2}{3}k^2u_{\textsc{cn}}=0,\qquad\implies\qquad u_{\textsc{cn}}=\alpha_k\mathrm{e}^{-i\sqrt{\frac{2}{3}}k\eta}+\beta_k\mathrm{e}^{i\sqrt{\frac{2}{3}}k\eta}.
\end{equation}
Taking the normalization above as the canonical one, we can impose the Bunch--Davies vacuum, which selects only positive frequencies, and thus gives
\begin{equation}\label{eq|eqinicondmuk}
    \alpha_k=\left(\sqrt{\frac{2}{3}}k\right)^{-1/2},\qquad\beta_k=0.
\end{equation}
Using the sub-horizon limit of the comoving curvature perturbation~\eqref{eq|comosubhnmc}, the sub-horizon limit of definition~\eqref{eq|mukvarcn}, and the solution for the velocity field~\eqref{eq|nmcsolvvd}, we obtain the corresponding initial conditions for $\mathcal{R}$:
\begin{equation}\label{eq|eqinicond}
    C_{1,\textsc{nmc}}\sim \frac{10^{17}}{k^2}\beta_k=0,\qquad C_{2,\textsc{nmc}}\sim \frac{10^{17}}{k^2}\alpha_k\sim10^{17}\frac{1}{k^{5/2}}.
\end{equation}
When perturbations exit the horizon, the growing mode in~\eqref{eq|comosuphnmc} rapidly dominates over the constant one, and by the end of the NMC-dominated epoch their amplitude is
\begin{equation}\label{eq|comoevo}
    \mathcal{R}\sim k^2C_{2,\textsc{nmc}}\eta_{\textsc{nmc}}^{-1}\sim\frac{10^{17}}{\eta_{\textsc{nmc}}}\frac{1}{k^{1/2}},
\end{equation}
as illustrated in figure~\ref{fig|inflation}.

At later times the evolution follows GR, where the constant mode dominates, so that perturbations re-enter the horizon with amplitude~\eqref{eq|comoevo}. The corresponding power spectrum at horizon entry is
\begin{equation}\label{eq|comopowerspectrum}
    \mathcal{P}_{\mathcal{R}}(k)= \frac{k^3}{(2\pi)^2}\left|\mathcal{R}\right|^2\sim \frac{10^{32}}{\eta_{\textsc{nmc}}^2} k^2.
\end{equation}
The scalar spectral index is therefore $n_{\rm s}=3$, in clear tension with observations. Perturbations generated during the accelerated phase cannot account for the observed cosmological fluctuations, and alternative scenarios must be considered.

\subsection{Power spectrum within bouncing scenario}\label{subsec|bounce}

An alternative to generating primordial fluctuations during the accelerated phase is that they originate during a contracting phase in a bouncing cosmology. In standard models of primordial perturbations in bouncing cosmologies an asymmetric bounce is normally required. Indeed, in order to provide the observed nearly scale-invariant spectrum, the pre-bounce Universe has to be matter dominated while becoming radiation dominated immediately after the bounce. This is because, perturbations generated deep inside the horizon are normalized by imposing the Bunch--Davies vacuum, $u\propto k^{-1/2}$. As they exit the horizon during contraction, the time-dependent mode grows and dominates over the constant one, leading to an amplification by a factor $k^{-1}$ and hence to a scale-invariant power spectrum at horizon re-entry after the bounce~\cite{Finelli:2001sr,Brandenberger:2012zb}.

Interestingly, in our case, no change in the matter content of the pre-bounce Universe is required, since the model itself predicts a symmetric cosmological bounce for negative spatial curvature. Nonetheless, as we will show, obtaining an almost scale-invariant spectrum requires a different (albeit we shall argue not unnatural) tuning of initial conditions with respect to the Bunch-Davies vacuum.

In order to make the above discussion quantitative, let us introduce the aforementioned Mukhanov variable $u=a\mathcal{R}$.
As anticipated, by doing so the first equation in~\eqref{eq|eqcomogr} takes the form of a Klein--Gordon equation for modes of wavenumber $\tilde{k}=c_sk$ with a time-dependent effective mass. Explicitly
\begin{equation}
    u''+\left(c_s^2k^2-\frac{1}{2}\left(1-3w\right)\mathcal{H}^2\right)u=0\,.
\end{equation}
To study the quantum regime, we consider the deep sub-horizon limit, in which the equation simplifies and admits the solution
\begin{equation}\label{eq|muksolbounce}
    u''+c_s^2k^2u=0\qquad \implies\qquad u=\alpha_k\mathrm{e}^{-ic_sk\eta}+\beta_k\mathrm{e}^{ic_sk\eta}.
\end{equation}

Now, it is worth noticing that, during its pre-bounce contracting phase, the Universe was probably already filled by matter and radiation, so that the Bunch-Davies vacuum may not be its more appropriate description. While it is quite difficult to ``guess'' an alternative, non-vacuum, initial condition, we can still ``reverse-engineer'' checking which ones would have lead in our bouncing scenario to the observed quasi scale-invariant spectrum.

In order to construct suitable initial conditions, we recall that standard commutation relations imply
\begin{equation}\label{eq|commutation}
    u\bar{u}'-u'\bar{u}=2i
    \qquad\implies\qquad
    \left|\alpha_k\right|^2-\left|\beta_k\right|^2=\frac{1}{c_s k}.
\end{equation}
Comparing the sub-horizon solution for the comoving curvature perturbation in equation~\eqref{eq|comosubhgr} with the solution for the Mukhanov variable in equation~\eqref{eq|muksolbounce}, this condition translates into the following constraint on the integration constants of $\mathcal{R}$:
\begin{equation}\label{eq|eqinicondbounce}
    \left|C_2\right|^2-\left|C_1\right|^2=\frac{\pi}{2}.
\end{equation}
This relation fixes one combination of the two integration constants, leaving a single free parameter once the coefficients are taken to be real. Rather than working directly with $C_1$ and $C_2$, it is convenient to parametrize the remaining freedom by introducing a characteristic scale in momentum space, $\bar{k}^{(\textsc{s})}$, defined as the wavenumber for which the amplitudes of the constant and growing modes are equal at $\eta=\bar{\eta}$, which corresponds to
\begin{equation}\label{eq|inconscale}
  \frac{\left|C_1+C_2\right|}{\left|C_1-C_2\right|}=\left|\frac{\Gamma\left(1-\nu\right)\Gamma\left(-\nu\right)}{2^{2\nu}\pi}\left(c_s\bar{k}^{(\textsc{s})}\bar{\eta}\right)^{2\nu}+\cot\left(\pi\nu\right)\right|.  
\end{equation}
With this definition, modes with $k<\bar{k}^{(\textsc{s})}$ reach $\eta=\bar{\eta}$ dominated by the growing mode, whereas modes with $k>\bar{k}^{(\textsc{s})}$ dominated by the constant mode.

Combining equations~\eqref{eq|eqinicondbounce} and~\eqref{eq|inconscale}, the integration constants can be written as
\begin{equation}\label{eq|initcond}
\begin{split}
    C_1=&\ \mp\sqrt{\frac{\pi}{8}}\frac{1-\left|\frac{\Gamma\left(1-\nu\right)\Gamma\left(-\nu\right)}{2^{2\nu}\pi}\left(c_s\bar{k}^{(\textsc{s})}\bar{\eta}\right)^{2\nu}+\cot\left(\pi\nu\right)\right|}{\left|\frac{\Gamma\left(1-\nu\right)\Gamma\left(-\nu\right)}{2^{2\nu}\pi}\left(c_s\bar{k}^{(\textsc{s})}\bar{\eta}\right)^{2\nu}+\cot\left(\pi\nu\right)\right|^{1/2}},\\
    C_2=&\ \pm\sqrt{\frac{\pi}{8}}\frac{1+\left|\frac{\Gamma\left(1-\nu\right)\Gamma\left(-\nu\right)}{2^{2\nu}\pi}\left(c_s\bar{k}^{(\textsc{s})}\bar{\eta}\right)^{2\nu}+\cot\left(\pi\nu\right)\right|}{\left|\frac{\Gamma\left(1-\nu\right)\Gamma\left(-\nu\right)}{2^{2\nu}\pi}\left(c_s\bar{k}^{(\textsc{s})}\bar{\eta}\right)^{2\nu}+\cot\left(\pi\nu\right)\right|^{1/2}}.
\end{split}
\end{equation}
These conditions correspond to an initial state of the Mukhanov variable with occupation number
\begin{equation}\label{eq|occnumb}
    n_k=\left|\beta_k\right|^2=\frac{\left|C_1\right|^2}{\frac{\pi}{2}c_sk}=\frac{\left(1-\left|\frac{\Gamma\left(1-\nu\right)\Gamma\left(-\nu\right)}{2^{2\nu}\pi}\left(c_s\bar{k}^{(\textsc{s})}\bar{\eta}\right)^{2\nu}+\cot\left(\pi\nu\right)\right|\right)^2}{4\left|\frac{\Gamma\left(1-\nu\right)\Gamma\left(-\nu\right)}{2^{2\nu}\pi}\left(c_s\bar{k}^{(\textsc{s})}\bar{\eta}\right)^{2\nu}+\cot\left(\pi\nu\right)\right|c_s k}.
\end{equation}
The Bunch--Davies vacuum is recovered for the particular choice
\begin{equation}
    \bar{k}^{(\textsc{s})}_{\textsc{bd}}=\frac{2}{c_s\bar{\eta}}\left(\frac{\pi\left(\pm1-\cot\left(\pi\nu\right)\right)}{\Gamma(1-\nu)\Gamma(-\nu)}\right)^{1/2\nu},
\end{equation}
for which $C_1=0$ and $C_2=\sqrt{\pi/2}$, and therefore $n_k=0$. Values of $\bar{k}^{(\textsc{s})}$ either smaller or larger than $\bar{k}^{(\textsc{s})}_{\textsc{bd}}$ correspond to excited states.

We note that adopting analogous initial conditions in the inflationary scenario, so as to keep the constant mode of equation~\eqref{eq|comosuphnmc} at horizon exit, would introduce an additional factor of $k$ in the comoving curvature perturbation. The resulting scalar spectral index would then be $n_{\rm s}=5$, making the predictions even more incompatible with observations.

\begin{figure}[t!]
    \centering
    \includegraphics[width=\textwidth]{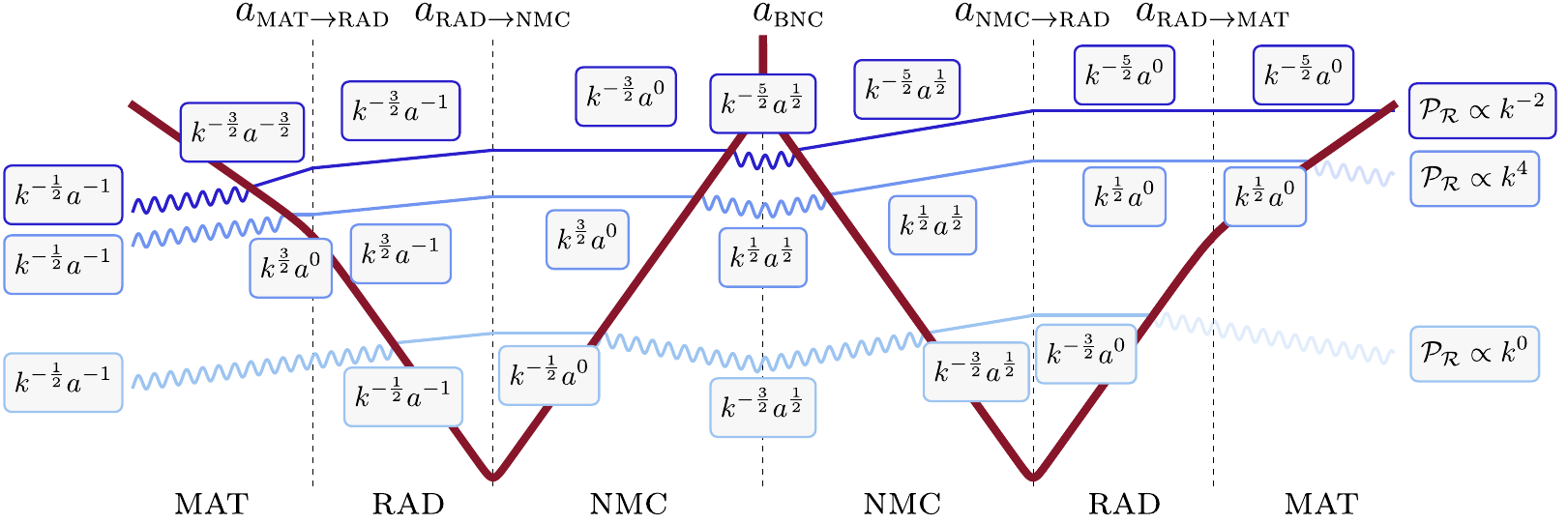}
    \caption{Pictorial representation of the evolution of perturbations generated during a contracting phase in a bouncing cosmology, showing their dependence on the wavenumber $k$ and on the scale factor $a$. All modes originate from a quantum state normalized as $k^{-1/2}$. The upper mode in the figure exits the horizon during matter domination and reaches matter-radiation equality with the growing mode dominant. The central mode also exits the horizon during matter domination, but reaches matter-radiation equality with the constant mode dominant. The lower mode exits the horizon during radiation domination. The right-hand side of the figure shows the resulting $k$-dependence of the power spectrum. 
    }
    \label{fig|bounce}
\end{figure}
With this freedom in the choice of initial conditions, we can now follow the evolution of perturbations up to the CMB and derive a prediction for the primordial power spectrum. However, as shown in figure~\ref{fig|bounce}, the evolution of perturbations in the bouncing scenario is considerably more involved than in the inflationary case. It is convenient to separate the evolution into different stages and discuss each of them independently.\\

\paragraph{Contraction phase during matter domination.} The evolution of perturbations in a matter-dominated Universe requires some care, since the solution in equation~\eqref{eq|solcomogr} becomes singular in the limit $c_s\to 0$, or identically constant if one sets $C_{1}=-C_{2}$ 
to avoid the divergence. A useful approximation is obtained by taking the limit $\omega\to 0$, while keeping $c_s$ finite. The full solution then reads
\begin{equation}\label{eq|solcomogrmat}
    \mathcal{R}=-C_{1,\textsc{mat}}\sqrt{\frac{2}{\pi c_s^3k^3}}\left(1-ic_sk\eta\right)\frac{\mathrm{e}^{ic_sk\eta}}{\eta^3}-C_{2,\textsc{mat}}\sqrt{\frac{2}{\pi c_s^3k^3}}\left(1+ic_sk\eta\right)\frac{\mathrm{e}^{-ic_sk\eta}}{\eta^3},
\end{equation}
where, during matter domination, $a\propto \eta^2$, so that $|\eta|\propto \sqrt{a}$ up to a normalization fixed by the background solution, and we made explicit that the integration constants $C_{1,\textsc{mat}}$ and $C_{2,\textsc{mat}}$ refer to the matter-dominated epoch. In the super-horizon limit it reduces to
\begin{equation}\label{eq|solcomogrmatsuph}
    \mathcal{R}_{k\ll\mathcal{H}}\sim \frac{i}{3}\sqrt{\frac{2}{\pi}}(c_sk)^{3/2}(C_{2,\textsc{mat}}-C_{1,\textsc{mat}})-\sqrt{\frac{2}{\pi}}\frac{C_{1,\textsc{mat}}+C_{2,\textsc{mat}}}{(c_sk)^{3/2}}a^{-3/2}.
\end{equation}

We choose as initial conditions
\begin{equation}\label{eq|inicomobouncem}
    C_{1,\textsc{mat}}=\mp\sqrt{\frac{\pi}{8}}\frac{1-\left(\frac{c_s}{\sqrt[3]{3}}\bar{k}^{(\textsc{s})}_{\textsc{mat}} \sqrt{a_{\textsc{mre}}}\right)^3}{\left(\frac{c_s}{\sqrt[3]{3}}\bar{k}^{(\textsc{s})}_{\textsc{mat}} \sqrt{a_{\textsc{mre}}}\right)^{3/2}},\qquad\quad C_{2,\textsc{mat}}=\pm\sqrt{\frac{\pi}{8}}\frac{1+\left(\frac{c_s}{\sqrt[3]{3}}\bar{k}^{(\textsc{s})}_{\textsc{mat}} \sqrt{a_{\textsc{mre}}}\right)^3}{\left(\frac{c_s}{\sqrt[3]{3}}\bar{k}^{(\textsc{s})}_{\textsc{mat}} \sqrt{a_{\textsc{mre}}}\right)^{3/2}},
\end{equation}
so that modes with $k<\bar{k}^{(\textsc{s})}_{\textsc{mat}} $ reach matter–radiation equality dominated by the growing mode, while modes with $k>\bar{k}^{(\textsc{s})}_{\textsc{mat}} $ dominated by the constant mode.
Allowing for such an intermediate scale $\bar{k}^{(\textsc{s})}_{\textsc{mat}} $, comoving curvature perturbations at the end of this stage scale as
\begin{equation}\label{eq|comonmcmr}
    \mathcal{R}_{{\textsc{mat}}\to {\textsc{rad}}}\propto\begin{cases}
        k^{-3/2} & \text{for }\quad k<\bar{k}^{(\textsc{s})}_{\textsc{mat}} \\
        k^{3/2} & \text{for }\quad \bar{k}^{(\textsc{s})}_{\textsc{mat}} <k<k_{\textsc{mre}},
    \end{cases}
\end{equation}
where $k_{\textsc{mre}}\sim \mathcal{H}_{\textsc{mre}}\sim 1.04\times 10^{-2}\,\text{Mpc}^{-1}$, is the perturbation's wavenumber at matter-radiation equality.     

\paragraph{Contraction phase during radiation domination.} The evolution of perturbations in a radiation-dominated Universe is particularly simple, since the full solution in equation~\eqref{eq|solcomogr}, with $c_s^2\sim\omega\sim1/3$, reduces to
\begin{equation}\label{eq|solcomogrrad}
    \mathcal{R}= \frac{C_{1,\textsc{rad}}}{\sqrt{\frac{\pi}{2}\frac{1}{\sqrt{{3}}}k}}\frac{\mathrm{e}^{i\frac{1}{3}k\eta}}{\eta}+\frac{C_{2,\textsc{rad}}}{\sqrt{\frac{\pi}{2}\frac{1}{\sqrt{{3}}}k}}\frac{\mathrm{e}^{-i\frac{1}{3}k\eta}}{\eta},
\end{equation}
where, during radiation domination, $a\sim \eta$ up to a normalization fixed by the background solution. In the super-horizon limit it reduces to
\begin{equation}\label{eq|solcomogrradsuph}
    \mathcal{R}_{k\ll\mathcal{H}}\sim i\sqrt{\frac{2}{\pi}}\sqrt{\frac{1}{\sqrt{3}}k}\left(C_{2,\textsc{rad}}-C_{1,\textsc{rad}}\right)+\sqrt{\frac{2}{\pi}}\frac{C_{1,\textsc{rad}}+C_{2,\textsc{rad}}}{\sqrt{\frac{1}{\sqrt{3}}k}}a^{-1}.
\end{equation}
Modes that are already super-horizon are rescaled by the common factor $\eta_{\textsc{mre}}/\eta_{\textsc{nmc}}=a_{\textsc{mre}}/a_{\textsc{nmc}}$.
For modes exiting the horizon during this phase, we fix the integration constants as
\begin{equation}\label{eq|inicomobouncer}
    C_{1,{\textsc{rad}}}\sim\mp\sqrt{\frac{\pi}{8}}\frac{1-\frac{1}{\sqrt{3}}\bar{k}^{(\textsc{s})}_{\textsc{rad}} a_{\textsc{nmc}}}{\left(\frac{1}{\sqrt{3}}\bar{k}^{(\textsc{s})}_{\textsc{rad}} a_{\textsc{nmc}}\right)^{1/2}},\qquad\quad C_{2,\textsc{rad}}\sim\pm\sqrt{\frac{\pi}{8}}\frac{1+\frac{1}{\sqrt{3}}\bar{k}^{(\textsc{s})}_{\textsc{rad}} a_{\textsc{nmc}}}{\left(\frac{1}{\sqrt{3}}\bar{k}^{(\textsc{s})}_{\textsc{rad}} a_{\textsc{nmc}}\right)^{1/2}},
\end{equation}
so that perturbations with $k<\bar{k}^{(\textsc{s})}_{\textsc{rad}}$ reach the onset of the non-minimal coupling dominated by the growing mode, while those with $k>\bar{k}^{(\textsc{s})}_{\textsc{rad}}$ dominated by the constant mode. To remain consistent with observational constraints, however, only the growing modes should be observed, and thus we choose $\bar{k}^{(\textsc{s})}_{\textsc{rad}}>\max{\left(k_{\rm obs}\right)}$. 
These modes do not acquire additional $k$-dependence since they scale as $\eta^{-1}=a^{-1}$ both inside and outside the horizon. At the end of this stage, comoving curvature perturbations behave as
\begin{equation}\label{eq|comonmcrnmc}
    \mathcal{R}_{{\textsc{rad}}\to\textsc{nmc}}\propto\begin{cases}
        k^{-3/2} & \text{for }\quad k<\bar{k}^{(\textsc{s})}_{\textsc{mat}} \\
        k^{3/2} & \text{for }\quad \bar{k}^{(\textsc{s})}_{\textsc{mat}} <k<k_{\textsc{mre}}\\
        k^{-1/2} & \text{for }\quad k>k_{\textsc{mre}}.
    \end{cases}
\end{equation}

\paragraph{Contraction phase during NMC domination.} When the Universe reaches the NMC-dominated phase, all relevant modes are already super-horizon. Their evolution is described by equation~\eqref{eq|comosuphnmc}, which contains a constant mode and a time-dependent one scaling as $\eta^{-1}=a^{1/2}$. The latter rapidly decreases and becomes subdominant. Comparing the super-horizon behavior in equation~\eqref{eq|comosuphnmc} with the sub-horizon one in equation~\eqref{eq|comosubhnmc}, one finds that perturbations are amplified by a factor $k^{-1}$. At the bounce, comoving curvature perturbations therefore scale as
\begin{equation}\label{eq|comonmcbounce}
    \mathcal{R}_{\textsc{bnc}}\propto\begin{cases}
        k^{-5/2} & \text{for }\quad k<\bar{k}^{(\textsc{s})}_{\textsc{mat}} \\
        k^{1/2} & \text{for }\quad \bar{k}^{(\textsc{s})}_{\textsc{mat}} <k<k_{\textsc{mre}}\\
        k^{-3/2} & \text{for }\quad k>k_{\textsc{mre}}.
    \end{cases}
\end{equation}

\paragraph{Expansion phase during NMC domination.} In the sub-horizon regime, perturbations evolve as in equation~\eqref{eq|comosubhnmc}, namely as oscillations modulated by a factor $\eta^{-1}=a^{1/2}$. In a symmetric bounce scenario, each mode exits the horizon after the bounce at the same value of the scale factor at which it entered before the bounce, with a small common enhancement, as shown in Appendix~\ref{app|bounce}. No additional $k$-dependence is generated at horizon exit, since the time-dependent mode dominates due to its scaling as $\eta^{-1}=a^{1/2}$. Comoving curvature perturbations therefore reach the end of the NMC-dominated stage with
\begin{equation}\label{eq|comonmcnmcr}
    \mathcal{R}_{{\textsc{nmc}}\to {\textsc{rad}}}\propto\begin{cases}
        k^{-5/2} & \text{for }\quad k<\bar{k}^{(\textsc{s})}_{\textsc{mat}} \\
        k^{1/2} & \text{for }\quad \bar{k}^{(\textsc{s})}_{\textsc{mat}} <k<k_{\textsc{mre}}\\
        k^{-3/2} & \text{for }\quad k>k_{\textsc{mre}}.
    \end{cases}
\end{equation}

\paragraph{Expansion phase during radiation and matter dominations.} Finally, once the NMC becomes negligible, the evolution is again described by GR. In an expanding Universe, the constant mode of super-horizon perturbations dominates during both radiation and matter domination, so comoving curvature perturbations re-enter the horizon with the amplitude shown in equation~\eqref{eq|comonmcnmcr}.
Evaluating the power spectrum at horizon entry then gives
\begin{equation}\label{eq|powerspec}
    \left.\mathcal{P}_{\mathcal{R}}\right|_{\rm{H-entry}}\propto\begin{cases}
        k^{-2} & \text{for }\quad k<\bar{k}^{(\textsc{s})}_{\textsc{mat}} \\
        k^{4} & \text{for }\quad \bar{k}^{(\textsc{s})}_{\textsc{mat}} <k<k_{\textsc{mre}}\\
        k^{0} & \text{for }\quad k>k_{\textsc{mre}}.
    \end{cases}
\end{equation}
This clarifies the role of the intermediate scale $\bar{k}^{(\textsc{s})}_{\textsc{mat}} $. 
If we keep only the growing modes, as is usually done in cosmology, the $k^{-2}$ branch, combined with the integrated Sachs--Wolfe effect, would produce an excessive enhancement of power at large scales, in tension with observations. The $k^{4}$ branch, on the other hand, is partially compensated by the same effect.

While the piecewise approximation used is a strong simplification of cosmic history, it highlights a possible effect of a non-symmetric bounce. If matter-radiation equality in the contracting phase occurred earlier, corresponding to a longer radiation-dominated epoch during contraction than during expansion, the scale-invariant part of the power spectrum would be pushed towards larger scales. If matter-radiation equality in the contracting phase occurred later, corresponding to a scenario more similar to matter-bounce cosmologies, the $k^4$ dependence would extend towards smaller scales, making the power spectrum increasingly incompatible with observations.

Before confronting the resulting scalar spectrum with CMB temperature anisotropies, it is useful to consider the tensor sector. In particular, the observational bound on the tensor-to-scalar ratio already provides a constraint on $\bar{k}^{(\textsc{s})}_{\textsc{mat}}$.

\paragraph{Tensor-to-scalar ratio.} The evolution of tensor perturbations in the GR limit is nearly identical to that of comoving curvature perturbations, with the only difference that tensor modes propagate at the speed of light, $c_s=1$. During the NMC-dominated phase of the contracting Universe, however, the solution contains an exponentially growing mode. To preserve the validity of the linear approximation, this mode must be removed by setting $C^{+,\times}_{2,\textsc{nmc}}=0$ in equation~\eqref{eq|nmcT}. The remaining mode is exponentially suppressed during contraction, so that tensor perturbations reach the bounce with negligible amplitude. After the bounce, the same solution becomes exponentially growing. In a symmetric bounce scenario, this growth compensates for the suppression experienced during the contracting phase, and tensor perturbations therefore emerge from the NMC-dominated epoch with approximately the same amplitude they had when entering it. Unlike comoving curvature perturbations, tensor modes do not acquire the additional $k^{-1}$ enhancement generated during the NMC phase. Their power spectrum is therefore obtained from that of $\mathcal{R}$ by setting $c_s=1$ and removing this extra factor:
\begin{equation}\label{eq|powspecT}
    \mathcal{P}_h\propto k^2\left.\mathcal{P}_{\mathcal{R}}\right|_{c_s=1}.
\end{equation}
We parametrize the initial conditions for tensor perturbations in analogy with scalar perturbations in equations~\eqref{eq|inicomobouncem} and~\eqref{eq|inicomobouncer}, introducing the corresponding intermediate scales $\bar{k}^{(\textsc{t})}_{\textsc{mat}}$ and $\bar{k}^{(\textsc{t})}_{\textsc{rad}}$. The tensor-to-scalar ratio then takes the form 
\begin{equation}\label{eq|tentosca}
    r_{\textsc{mat}}= k^2\left(\frac{\bar{k}^{(\textsc{s})}_{\textsc{mat}}}{\bar{k}^{(\textsc{t})}_{\textsc{mat}}}\right)^3\frac{k^6+\left(\bar{k}^{(\textsc{t})}_{\textsc{mat}}\right)^6}{k^6+\left(\bar{k}^{(\textsc{s})}_{\textsc{mat}}\right)^6},\qquad r_{\textsc{rad}}= k^2\frac{\bar{k}^{(\textsc{s})}_{\textsc{rad}}}{\bar{k}^{(\textsc{t})}_{\textsc{rad}}}\frac{k^2+\left(\bar{k}^{(\textsc{t})}_{\textsc{rad}}\right)^2}{k^2+\left(\bar{k}^{(\textsc{s})}_{\textsc{rad}}\right)^2},
\end{equation}
during matter and radiation domination, respectively, and where $k$ is in units of $\mathcal{H}_0$. \\

\begin{figure}[t!]
    \centering
    \includegraphics[width=0.92\textwidth]{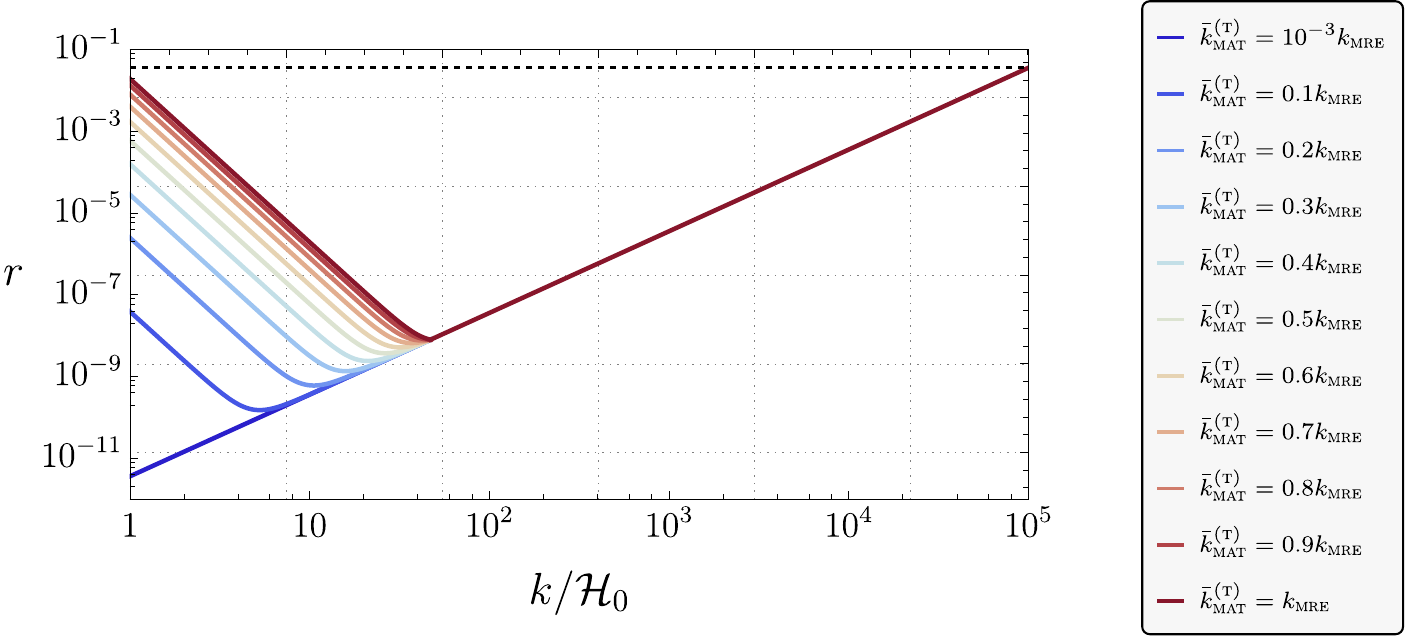}
    \caption{Possible shapes of the tensor-to-scalar ratio for different values of $\bar{k}^{(\textsc{t})}_{\textsc{mat}}$, given in the legend as fractions of $k_{\textsc{mre}}$. The parameter $\bar{k}^{(\textsc{t})}_{\textsc{rad}}$ is fixed to $\max\left(k_{\rm obs}\right)=10^5$, while $\bar{k}^{(\textsc{s})}_{\textsc{mat}}$ and $\bar{k}^{(\textsc{s})}_{\textsc{rad}}$ are chosen to satisfy continuity and to make $r$ saturate the Planck bound $r=0.036$ at $k=\max\left(k_{\rm obs}\right)$, under the conditions $\bar{k}^{(\textsc{s})}_{\textsc{mat}}<\min\left(k_{\rm obs}\right)=1$ and $\bar{k}^{(\textsc{s})}_{\textsc{rad}}>\max\left(k_{\rm obs}\right)=10^5$.}
    \label{fig|tensosca}
\end{figure}

At $k= \bar{k}^{(\textsc{s})}_{\textsc{mat}}$, the tensor-to-scalar ratio satisfies the Planck bound $r<r_{\textsc{pl}}=0.036$ only if $\bar{k}^{(\textsc{s})}_{\textsc{mat}}<\sqrt{r_{\textsc{pl}}}\sim0.190$, which is smaller than the wavenumber of the largest observable mode, $\min(k_{\rm obs})\sim1$ in these units. Since we also require $\bar{k}^{(\textsc{s})}_{\textsc{rad}}>\max(k_{\rm obs})$, the tensor-to-scalar ratio scales as
\begin{equation}\label{eq|tentoscasca}
    \begin{array}{c}
         r_{\textsc{mat}}\propto\begin{cases}
             k^{-4},& \mathrm{for}\ k<\bar{k}^{(\textsc{t})}_{\textsc{mat}}\\
             k^2,& \mathrm{for}\ \bar{k}^{(\textsc{t})}_{\textsc{mat}}<k<k_{\textsc{mre}}
         \end{cases},
    \end{array}
    \qquad\qquad\quad
    \begin{array}{c}
         r_{\textsc{rad}}\propto\begin{cases}
             k^2,& \mathrm{for}\ k_{\textsc{mre}}<k<\bar{k}^{(\textsc{t})}_{\textsc{rad}}\\
             k^4,& \mathrm{for}\ k>\bar{k}^{(\textsc{t})}_{\textsc{rad}}
         \end{cases}.
    \end{array}
\end{equation}

Requiring that the tensor-to-scalar ratio satisfy the Planck bound over the whole observable range places only weak constraints on $\bar{k}^{(\textsc{t})}_{\textsc{mat}}$ and $\bar{k}^{(\textsc{t})}_{\textsc{rad}}$. As a result, the precise shape of the tensor-to-scalar ratio is not a robust prediction of the model. Nevertheless, if $\bar{k}^{(\textsc{t})}_{\textsc{rad}}$ is chosen such that $r$ saturates the observational bound at the largest observable scales, and continuity is imposed, the tensor-to-scalar ratio as a function of the wavenumber $k$ resembles the examples shown in figure~\ref{fig|tensosca}: in the radiation-dominated branch, $r$ always increases with $k$, while in the matter-dominated branch it can either increase—--leading to very small values on large scales—--or decrease, making it potentially accessible to future observations.

A tensor-to-scalar ratio that could be detected by future experiments, while remaining close to the current upper bound on large scales, requires $\bar{k}^{(\textsc{s})}_{\textsc{mat}}<\sqrt{r_{\textsc{pl}}}$, a larger $\bar{k}^{(\textsc{t})}_{\textsc{mat}}\sim k_{\textsc{mre}}$, and large $\bar{k}^{(\textsc{s})}_{\textsc{rad}},\ \bar{k}^{(\textsc{t})}_{\textsc{rad}}>\max(k_{\rm obs})$. This implies that, at least during matter-domination, perturbations start in highly excited states, with scalar perturbations having a larger occupation number than tensor perturbations\footnote{We note that the difference between the scalar and tensor occupation numbers is linked to the requirement of having a tensor-to-scalar ratio detectable by future experiments, and is not a prediction of the model.}.

\begin{figure}[t!]
    \centering
    \includegraphics[width=0.8\textwidth]{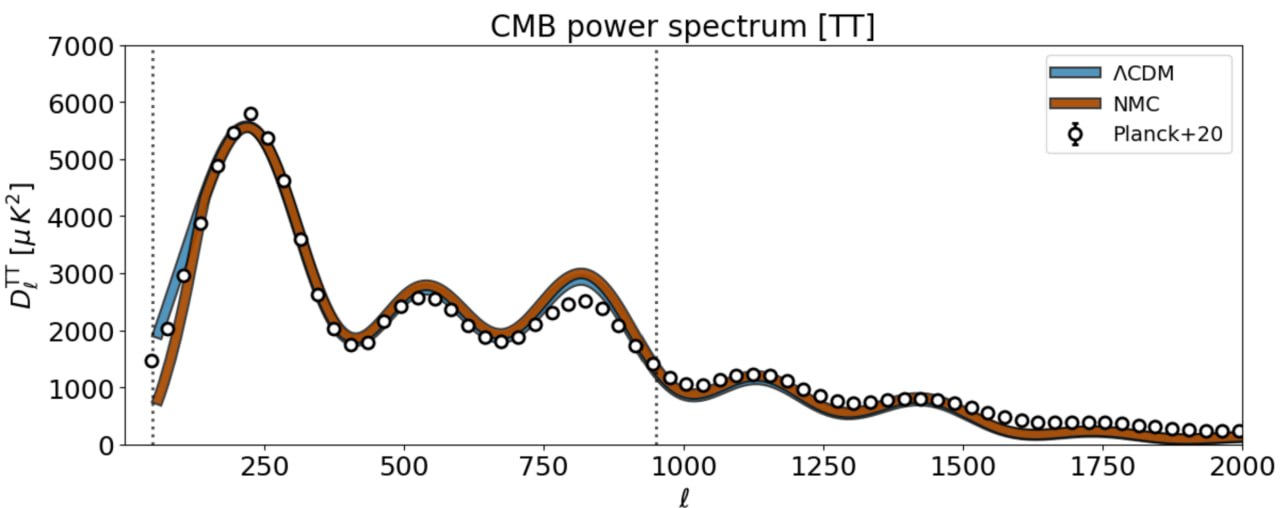}
    \caption{Fit of Planck data with our prediction for the primordial power spectrum in equation~\eqref{eq|powerspec}.}
    \label{fig|powerspec}
\end{figure}

\paragraph{Fit of Planck TT-power spectrum.} We use the matter power spectrum in equation~\eqref{eq|powerspec} to compute the CMB temperature angular power spectrum using the Weinberg's \cite{Weinberg2008,Baumann2022} semi-analytic approximation for the Fourier coefficients. Note that such a formula is known to have same limitations with respect to the full numerical solutions of the Boltzmann equation for the photon-baryon fluid: it slightly underestimates the first peak, slightly overestimates the third peak, and slightly under-damp the high-$\ell$ tail; however, these deviations are rather small, limited to $\lesssim 10\%$ for the peak amplitudes and to $\lesssim 1\%$ for the peak positions. Since it is quite complex to implement in the present Boltzmann solvers (e.g., \texttt{CAMB} or \texttt{CLASS}) the peculiarities of the NMC evolution, and since the piece-wise powerlaw description of the input NMC matter power spectrum is anyhow very crude, for a first semi-quantitative assessment we can confidently rely on the Weinberg's approximation. Choosing a $\bar{k}^{(\textsc{s})}_{\textsc{mat}}<\min(k_{\rm obs})$, we obtain the spectrum shown in figure~\ref{fig|powerspec}.
Overall, the agreement of our result with the Planck \cite{Planck2018} data is encouraging, and could motivate further, more precise analysis.

\section{Discussion and conclusions}

In this paper, we studied the generation of primordial cosmological perturbations in a Universe where dark matter is non-minimally coupled to gravity at the fluid level. The model assumes a direct coupling between matter and curvature which is subdominant with respect to Standard Model interactions, so that at low energies it is expected to be relevant only for dark matter. At the practical level, we introduced in the action a covariant contraction of the Einstein tensor with the stress-energy tensor of pressureless dark matter, together with a characteristic length scale that sets the strength of the interaction.

At the background level, the non-minimal coupling drives a phase of accelerated expansion at early times, lasting long enough to solve the horizon and flatness problems. At even earlier times, for negative spatial curvature, the initial singularity can be avoided, since the Universe undergoes a bounce from a previous contracting phase.

In section~\ref{sec|theory}, we presented the full equations for scalar, vector, and tensor perturbations. In the scalar sector, we showed that the equations for comoving curvature perturbations can be solved analytically both in the GR limit and in the regime where the non-minimal interaction dominates, separately; in the latter case, comoving curvature perturbations grow as the Universe expands. In the vector sector, the equations can be solved analytically without any approximation, and we found that vector modes are in general strongly suppressed at late times. In the tensor sector, analytic solutions can again be obtained in both the GR limit and the NMC-dominated epoch; in the latter case, tensor perturbations are described by exponential modes, which either invalidate the linear approximation or remain strongly suppressed.

In section~\ref{sec|evolution}, we showed how perturbations may be generated from quantum fluctuations and how they evolve up to the present epoch. If primordial fluctuations are generated during the accelerated expansion phase, perturbations do not receive the $k^{-1}$ enhancement present in standard inflationary scenarios. The model therefore predicts a scalar spectral index $n_{\rm s}=3$, in clear tension with observations.

If instead perturbations are generated during the contracting phase of a bouncing cosmology, the resulting scalar power spectrum is scale-invariant for multipoles $\ell\gtrsim150$, while perturbations at small wavenumber are suppressed as $k^4$. The integrated Sachs–Wolfe effect partially compensates for this suppression, and the resulting CMB power spectrum shows encouraging agreement with Planck data. The tensor-to-scalar ratio can be expressed in terms of several free parameters, and the model is therefore not highly predictive in this respect. Nevertheless, it is straightforward to obtain values consistent with current constraints while still leaving room for a possible detection in future experiments.

Obtaining observables compatible with current data requires a different prescription for the initial conditions of the perturbations. In particular, while standard inflationary scenarios usually assume that the perturbations begin in a vacuum state, our model requires that they start in an excited state. While this assumption is motivated a posteriori by the requirement of reproducing a primordial power spectrum compatible with observations, it does not appear unnatural. In contrast with inflationary scenarios, the past history of the Universe before a contracting epoch in a bouncing cosmology could be complex and leave relic perturbations. In particular, mechanisms such as thermal fluctuations, interactions, nonlinear dynamics, or a previous cosmological cycle could generate or preserve a non-vacuum state of primordial perturbations.

It could be argued that the normalization $n_k\propto k^{-1}$ might not be the most appropriate form for such a state. However, it is interesting to note that the occupation number in equation~\eqref{eq|occnumb}, $n_k=2|C_1|^2/(\pi c_sk)$, has the same form as the low-energy, or Rayleigh--Jeans, limit of a Bose--Einstein distribution. Indeed, for modes with dispersion relation $\omega_k=c_sk$,
\begin{equation}
n_k=\frac{1}{\exp\left(\frac{\hbar c_sk}{k_bT}\right)-1}\sim \frac{k_bT}{\hbar c_sk},
\end{equation}
when $\hbar c_sk\ll k_bT$. Comparing this expression with equation~\eqref{eq|occnumb}, we can associate an effective temperature
\begin{equation}
T_{\rm{eff}}=\frac{2\hbar\mathcal{H}_0|C_1|^2}{\pi k_b}.
\end{equation}
Indeed, during matter-domination we imposed $\bar{k}^{(s)}_{\textsc{mat}}<\sqrt{r_{\textsc{pl}}}\approx 0.19$, which implies an effective temperature $T_{\rm{eff}}\gtrsim4\times10^{-10}$~K. On the other hand, even for the highest-energy observable modes, $\hbar c_s \max{(k_{\rm obs})}\approx 2\times10^{-51}$~J~$
\approx 2\times10^{-28} k_b$~K, using $c_s=10^{-4}$. Therefore,
\begin{equation}
\frac{\hbar c_s \max{(k_{\rm obs})}}{k_bT_{\rm{eff}}}\lesssim 5\times10^{-19}\ll1,
\end{equation}
so that the relevant modes are well within the Rayleigh--Jeans regime. Thus, over the range of observable modes considered here, a thermal Bose--Einstein distribution would have the same $k^{-1}$ dependence as the occupation number required by our model.

This comparison does not imply that the primordial perturbations were in thermal equilibrium. Rather, it shows that their required $k^{-1}$ occupation number is compatible with the low-energy behaviour of a thermal distribution and therefore does not appear to require an intrinsically exotic form of the initial state. Interestingly, the corresponding energy density carried by the perturbations is of order $\rho_{\rm{pert}}\sim 10^{-30}\,\text{Kg}/\text{m}^3$, which is negligible for the background evolution.

To conclude, our analysis of the power spectrum of primordial perturbations shows that the generation of such perturbations during the phase of accelerated expansion is ruled out by observations. By contrast, the generation of the primordial spectrum in the contracting phase of a bouncing cosmology is compatible with observational constraints. This suggests that a nonsingular cosmological evolution may be required in models in which matter has a non-trivial interaction with gravity. In the specific case of the model considered here, this also translates into the prediction of negative spatial curvature.

These results are encouraging, but several open issues remain. First, the present analysis relies on a piecewise approximation for the background evolution, which should be replaced by a full numerical treatment to obtain precise predictions. Second, to complete the analysis of perturbations in the accelerated expansion phase, an explicit expansion of the action to second order in perturbation theory is required; together with fixing the normalization of the initial state of the perturbations, this would also clarify the presence of ghost, tachyonic, or strong-coupling instabilities. Third, the physical origin of the excited initial state of the perturbations in the bouncing scenario deserves further study. Fourth, our treatment assumes a symmetric bounce and specific parameter choices, and it remains to be understood how robust the results are under more general dynamics. Finally, while the present work focused on the scalar spectral index and tensor-to-scalar ratio, further observational tests should include non-Gaussianity and large-scale structure constraints.

\acknowledgments

We express our sincere gratitude to Marco Bruni for illuminating discussions. This work was partially funded from the projects: `Data Science methods for MultiMessenger Astrophysics \& Multi-Survey Cosmology' funded by the Italian Ministry of University and Research, Programmazione triennale 2021/2023 (DM n.2503 dd. 9 December 2019), Programma Congiunto Scuole; Italian Research Center on High Performance Computing Big Data and Quantum Computing (ICSC), project funded by European Union - NextGenerationEU - and National Recovery and Resilience Plan (NRRP) - Mission 4 Component 2 within the activities of Spoke 3 (Astrophysics and Cosmos Observations);  European Union - NextGenerationEU under the PRIN MUR 2022 project n. 20224JR28W `Charting unexplored avenues in Dark Matter'; INAF GO-GTO Normal 2023 funding scheme with the project "Serendipitous H-ATLAS-fields Observations of Radio Extragalactic Sources (SHORES)".
AL has been supported by the Istituto Nazionale di Fisica Nucleare (INFN) via the specific national initiative QGSKY.

\appendix

\section{Solution of the scalar sector in the NMC-epoch}\label{app|nmccomoving}

While equations~\eqref{eq|eqcomonmc} are already much simpler than the full set of equations, they can acquire a very compact form by defining the variable
\begin{equation}\label{eq|nmcvariable}
    \tilde{\mathcal{R}}:=\mathcal{R}-\frac{\mathcal{H}}{4}v_{\textsc{dm}}.
\end{equation}
With this choice of variable, the equations~\eqref{eq|eqcomonmc} can be written as
\begin{equation}\label{eq|eqnmcsolvable}
    \begin{split}
        \tilde{\mathcal{R}}''-2\mathcal{H}\frac{2k^2-3\mathcal{H}^2}{4k^2-3\mathcal{H}^2}\tilde{\mathcal{R}}'+\frac{4}{3}\frac{2k^2-3\mathcal{H}^2}{4k^2-3\mathcal{H}^2}\tilde{\mathcal{R}}=&\ 0,\\
        \delta_{\textsc{dm}}'+6\mathcal{H}v_{\textsc{dm}}'+\left(k^2+3\mathcal{H}^2\right)v_{\textsc{dm}}-\frac{36\mathcal{H}^2}{4k^2-3\mathcal{H}^2}\tilde{\mathcal{R}}'+\frac{24k^2\mathcal{H}}{4k^2-3\mathcal{H}^2}\tilde{\mathcal{R}}=&\ 0,\\
        \frac{v_{\textsc{dm}}'}{k}+6\mathcal{H}\frac{v_{\textsc{dm}}}{k}+\delta_{\textsc{dm}}+\frac{4}{\mathcal{H}}\frac{2k^2-9\mathcal{H}^2}{k^2+3\mathcal{H}^2}\tilde{\mathcal{R}}'+\frac{6\left(2k^2+\mathcal{H}^2\right)}{4k^2-3\mathcal{H}^2}\tilde{\mathcal{R}}=&\ 0.
    \end{split}
\end{equation}
Since during the NMC epoch the comoving Hubble parameter scales as $\mathcal{H}=-2/\eta$, we can solve analytically the first of the equations~\eqref{eq|eqnmcsolvable}, yielding
\begin{equation}\label{eq|nmcsolvr}
    \tilde{\mathcal{R}}= \tilde{C}_{1,\textsc{nmc}}\left(1-i\sqrt{\frac{2}{3}}k\eta-\frac{2}{3}k^2\eta^2\right)\frac{\mathrm{e}^{i\sqrt{\frac{2}{3}}k\eta}}{\eta^3}+\tilde{C}_{2,\textsc{nmc}}\left(1+i\sqrt{\frac{2}{3}}k\eta-\frac{2}{3}k^2\eta^2\right)\frac{\mathrm{e}^{-i\sqrt{\frac{2}{3}}k\eta}}{\eta^3},
\end{equation}
which can be used in the other two equations~\eqref{eq|eqnmcsolvable} to find the velocity field and density perturbation
\begin{equation}\label{eq|nmcsolvvd}
\begin{split}
    v_{\textsc{dm}}=\, &\tilde{C}_{1,\textsc{nmc}}\frac{2}{5}\left(5-i5\sqrt{\frac{2}{3}}k\eta-\frac{2}{3}k^2\eta^2\right)\frac{\mathrm{e}^{i\sqrt{\frac{2}{3}}k\eta}}{\eta^2}+\tilde{C}_{2,\textsc{nmc}}\frac{2}{5}\left(5+i5\sqrt{\frac{2}{3}}k\eta-\frac{2}{3}k^2\eta^2\right)\frac{\mathrm{e}^{-i\sqrt{\frac{2}{3}}k\eta}}{\eta^2}+\\
    &+ \tilde{C}_{3,\textsc{nmc}}\mathrm{e}^{k\eta}+\tilde{C}_{4,\textsc{nmc}}\mathrm{e}^{-k\eta},\\[0.4cm]
    \delta_{\textsc{dm}}=\, & \tilde{C}_{1,\textsc{nmc}}\frac{12}{20}\left(20-i20\sqrt{\frac{2}{3}}k\eta-3\frac{2}{3}k^2\eta^2-i\left(\frac{2}{3}\right)^{3/2}k^3\eta^3\right)\frac{\mathrm{e}^{i\sqrt{\frac{2}{3}}k\eta}}{\eta^3}+\\
    &+\tilde{C}_{2,\textsc{nmc}}\left(20+i20\sqrt{\frac{2}{3}}k\eta-3\frac{2}{3}k^2\eta^2+i\left(\frac{2}{3}\right)^{3/2}k^3\eta^3\right)\frac{\mathrm{e}^{-i\sqrt{\frac{2}{3}}k\eta}}{\eta^3}+\\
    &+ \tilde{C}_{3,\textsc{nmc}}\left(12-k\eta\right)\frac{\mathrm{e}^{k\eta}}{\eta}+\tilde{C}_{4,\textsc{nmc}}\left(12+k\eta\right)\frac{\mathrm{e}^{-k\eta}}{\eta},
\end{split}
\end{equation}
where the $\tilde{C}_{i,\textsc{nmc}}$ are integration constants. The variable $\tilde{\mathcal{R}}$ and the velocity field $v_{\textsc{dm}}$ combine to give the simple form for the comoving curvature perturbation in equation~\eqref{eq|solcomonmc}, after the definition of free parameters $C_{1,\textsc{nmc}}:=-\frac{8}{15}\tilde{C}_{1,\textsc{nmc}}$, $C_{2,\textsc{nmc}}:=-\frac{8}{15}\tilde{C}_{2,\textsc{nmc}}$, $C_{3,\textsc{nmc}}:=-\frac{1}{2}\tilde{C}_{3,\textsc{nmc}}$ and $C_{4,\textsc{nmc}}:=-\frac{1}{2}\tilde{C}_{4,\textsc{nmc}}$.

\section{Evolution of perturbations around the bounce}\label{app|bounce}

Close to the bounce, the conformal Hubble parameter no longer satisfies the differential equation
\begin{equation}\label{eq|appH1}
    \mathcal{H}'\sim-\frac{1}{2}\left(1+3w\right)\mathcal{H}^2,
\end{equation}
but instead obeys
\begin{equation}\label{eq|appH2}
    \mathcal{H}'\sim\frac{1}{2}\left(\mathcal{H}^2-\kappa\right).
\end{equation}
For negative spatial curvature, its dependence on conformal time is therefore
\begin{equation}
    \mathcal{H}\sim\sqrt{|\kappa|}\tan{\left(\frac{1}{2}\sqrt{|\kappa|}\,\eta\right)},
\end{equation}
and the conformal time can be written in terms of the scale factor as
\begin{equation}
    \eta\sim\frac{2}{\sqrt{|\kappa|}}\arccos{\left(\sqrt{\frac{a_{\textsc{bnc}}}{a}}\right)},\qquad\implies\qquad\mathcal{H}\sim\sqrt{|\kappa|\left(\frac{a}{a_{\textsc{bnc}}}-1\right)},
\end{equation}
where $a_{\textsc{bnc}}$ is the value of the scale factor at the bounce~\eqref{eq|scalefacbounce}.

If we use equation~\eqref{eq|appH2} instead of equation~\eqref{eq|appH1}, the expression for the comoving curvature perturbation becomes
\begin{equation}
    \mathcal{R}=\Psi+\frac{2}{3}\frac{\Psi+\frac{\Psi'}{\mathcal{H}}-\chi\left(\Psi+\frac{\Psi'}{\mathcal{H}}-\left(\frac{3}{2}+\frac{|\kappa|}{2\mathcal{H}^2}\right)\mathcal{H}v_{\textsc{dm}}+2v_{\textsc{dm}}'\right)}{\frac{1}{3}-\frac{|\kappa|}{2\mathcal{H}^2}+\frac{4}{3}\chi\left(1+\frac{|\kappa|}{\mathcal{H}^2}\right)},
\end{equation}
and the equations for $\mathcal{R},\ \delta_{\textsc{dm}}$, and $v_{\textsc{dm}}$, in the limit of $\chi\gg 1$ take the form
\begin{equation}\label{eq|appeqbounce}
    \begin{split}
        \mathcal{R}''+\frac{\kappa  \left(27 \kappa +4 k^2\right)-\left(15 \kappa +4 k^2\right)\mathcal{H}^2+6 \mathcal{H}^4}{18 \kappa ^2+ \left(4 k^2-9 \kappa \right)\mathcal{H}^2-3 \mathcal{H}^4}\mathcal{H}\mathcal{R}'+&\\
        +\frac{9\kappa^2\left(4k^2-15\kappa\right)+4\left(4k^2-9\kappa\right)\left(k^2+3\kappa\right)\mathcal{H}^2-3\left(8k^2-9\kappa\right)\mathcal{H}^4}{6 \left(18\kappa^2+\left(4k^2-9\kappa\right)\mathcal{H}^2-3\mathcal{H}^4\right)}\mathcal{R}+\frac{5\kappa+\mathcal{H}^2}{4\left(\kappa-\mathcal{H}^2\right)}\mathcal{H}v_{\textsc{dm}}''+&\\
        -\frac{3\left(30\kappa^4-12\kappa^3\mathcal{H}^2+\kappa\left(16k^2-19\kappa\right)\mathcal{H}^4-10\kappa\mathcal{H}^6+\mathcal{H}^8\right)}{4\left(\kappa-\mathcal{H}^2\right)\left(18\kappa^2+\left(4k^2-9\kappa\right)\mathcal{H}^2-3\mathcal{H}^4\right)}v_{\textsc{dm}}'+&\\
        -\bigg(6\kappa^3\left(16k^2-153\kappa\right)-\kappa\left(21k^4+900k^2\kappa-261\kappa^2\right)\mathcal{H}^2-\left(16k^4-276k^2\kappa-927\kappa^2\right)\mathcal{H}^4+&\\
        +3\left(8k^2-87\kappa\right)\mathcal{H}^6+9\mathcal{H}^8\bigg)\bigg(24\left(\kappa-\mathcal{H}^2\right)\left(18\kappa^2+\left(4k^2-9\kappa\right)\mathcal{H}^2-3\mathcal{H}^4\right)\bigg)^{-1}\mathcal{H}v_{\textsc{dm}}+&\\
        +\frac{9\kappa^3+\kappa\left(4k^2+9\kappa\right)\mathcal{H}^2-\left(2k^2+9\kappa\right)\mathcal{H}^4+3\mathcal{H}^6}{\left(\kappa-\mathcal{H}^2\right)\left(18\kappa^2+\left(4k^2-9\kappa\right)\mathcal{H}^2-3\mathcal{H}^4\right)}\kappa\delta_{\textsc{dm}}&\,=0,\\[0.4cm]
        \delta_{\textsc{dm}}'-\frac{18\kappa\mathcal{H}\left(2\kappa-\mathcal{H}^2\right)}{18\kappa^2+\left(4k^2-9\kappa\right)\mathcal{H}^2-3\mathcal{H}^4}\delta_{\textsc{dm}}-\frac{36\left(2\kappa^2-3\kappa\mathcal{H}^2+\mathcal{H}^4\right)}{18\kappa^2+\left(4k^2-9\kappa\right)\mathcal{H}^2-3\mathcal{H}^4}\mathcal{R}'+&\\
        -\frac{12\left(2k^2-3\kappa\right)\left(k-\mathcal{H}^2\right)}{18\kappa^2+\left(4k^2-9\kappa\right)\mathcal{H}^2-3\mathcal{H}^4}\mathcal{H}\mathcal{R}+\frac{3\left(6\kappa^2+\left(8k^2-9\kappa\right)\mathcal{H}^2-3\mathcal{H}^4\right)}{18\kappa^2+\left(4k^2-9\kappa\right)\mathcal{H}^2-3\mathcal{H}^4}\mathcal{H}v_{\textsc{dm}}'+&\\
        +\frac{18\kappa^2\left(2k^2-\kappa\right)+\left(2k^2-27\kappa\right)\left(4k^2+3\kappa\right)\mathcal{H}^2+6\left(k^2+18\kappa\right)\mathcal{H}^4-9\mathcal{H}^6}{2\left(18\kappa^2+\left(4k^2-9\kappa\right)\mathcal{H}^2-3\mathcal{H}^4\right)}v_{\textsc{dm}}&\,=0,\\[0.4cm]
        \frac{\left(5\kappa+\mathcal{H}^2\right)\left(9\kappa^2+\left(k^2-15\kappa\right)\mathcal{H}^2+3\mathcal{H}^4\right)}{2\left(18\kappa^2+\left(4k^2-9\kappa\right)\mathcal{H}^2-3\mathcal{H}^4\right)}v_{\textsc{dm}}'+&\\
        +\frac{4\kappa^2\left(5k^2-24\kappa\right)-\kappa\left(76k^2-69\kappa\right)\mathcal{H}^2+2\left(10k^2+21\kappa\right)\mathcal{H}^4+16\mathcal{H}^6}{4\left(18\kappa^2+\left(4k^2-9\kappa\right)\mathcal{H}^2-3\mathcal{H}^4\right)}\mathcal{H}v_{\textsc{dm}}+&\\
        -\frac{\left(2k^2+21\kappa-9\mathcal{H}^2\right)\left(\kappa-\mathcal{H}^2\right)}{2\left(18\kappa^2+\left(4k^2-9\kappa\right)\mathcal{H}^2-3\mathcal{H}^4\right)}\mathcal{H}\mathcal{R}'+&\\
        +\frac{3\left(\kappa-\mathcal{H}^2\right)\left(\kappa\left(2k^2+9\kappa\right)-\left(2k^2-4\kappa\right)\mathcal{H}^2-\mathcal{H}^4\right)}{2\left(18\kappa^2+\left(4k^2-9\kappa\right)\mathcal{H}^2-3\mathcal{H}^4\right)}\mathcal{R}+&\\
        -\frac{18\kappa^3+\kappa\left(8k^2+15\kappa\right)\mathcal{H}^2-4\left(k^2+3\kappa\right)\mathcal{H}^4+3\mathcal{H}^6}{4\left(18\kappa^2+\left(4k^2-9\kappa\right)\mathcal{H}^2-3\mathcal{H}^4\right)}\delta_{\textsc{dm}}&\,=0.
    \end{split}
\end{equation}
Equations~\eqref{eq|appeqbounce} do not admit a simple analytic solution; however, an approximate solution can be found in the limit $\mathcal{H}\to 0$. In this regime, $\mathcal{R},\ \delta_{\textsc{dm}}$ and $v_{\textsc{dm}}$ are linear combinations of exponential functions whose exponents,
\begin{equation}
    \pm k\eta\sqrt{\frac{4}{15}\left(-4+\frac{9|\kappa|}{8k^2}\pm\sqrt{-14-\frac{351|\kappa}{4k^2}+\frac{2241\kappa^2}{64k^4}}\right)},
\end{equation}
always contain non-zero real and imaginary parts. Expanding the solutions around $\eta= 0$, we obtain
\begin{equation}
\begin{split}
        \mathcal{R}\sim&\ C_{1,\textsc{bnc}}+C_{2,\textsc{bnc}}\eta,\\
        \delta_{\textsc{dm}}\sim&\ C_{3,\textsc{bnc}}+\left(4C_{2,\textsc{bnc}}-\left(k^2+3|\kappa|\right)C_{4,\textsc{bnc}}\right)\eta,\\
        v_{\textsc{dm}}\sim&\ C_{4,\textsc{bnc}}+\left(\frac{C_{3,\textsc{bnc}}}{5}+\frac{\left(2k^2-9|\kappa|\right)C_{1,\textsc{bnc}}}{15|\kappa|}\right)\eta.
\end{split}
\end{equation}
If the constant mode dominates before the bounce, while the time-dependent mode dominates after the bounce, all modes of the comoving curvature perturbation receive a small enhancement proportional to
\begin{equation}
    C_{2,\textsc{bnc}}\eta_{\textsc{bnc}\to \textsc{nmc}}\sim C_{2,\textsc{bnc}}\sqrt{\frac{4}{|\kappa|}\left(\frac{a_{\textsc{bnc} \to \textsc{nmc}}}{a_{\textsc{bnc}}}-1\right)}\sim C_{2,\textsc{bnc}}\sqrt{\frac{4}{|\kappa|}\frac{a_{\textsc{bnc}\to \textsc{nmc}}}{a_{\textsc{bnc}}}},
\end{equation} 
where $a_{\textsc{bnc} \to \textsc{nmc}}$ indicates the value of the scale factor for which the evolution has reduced to the standard NMC domination. This enhancement occurs because, although the conformal Hubble parameter vanishes at the bounce, the relevant long-wavelength condition is $k\eta\ll1$. Since the particle horizon
\begin{equation}
    d_{\textsc{h}}:=\int\mathrm{d}t\frac{1}{a(t)}=\eta,
\end{equation}
also vanishes at the bounce, all perturbation modes satisfy the long-wavelength condition and behave as effectively super-horizon modes in the sense of $kd_{\rm H}\ll1$, even though $k/\mathcal{H}$ diverges. The bounce therefore introduces only a small correction to the amplitude and does not modify the spectral dependence.

\bibliographystyle{unsrt}
\bibliography{biblio}
\end{document}